\documentclass[%
 reprint,
 amsmath,amssymb,
 aps,
floatfix,
]{revtex4-1}

\usepackage{graphicx}% Include figure files
\usepackage{dcolumn}% Align table columns on decimal point
\usepackage{bm}% bold math
\usepackage{slashed}
\begin{document}

\preprint{APS/123-QED}

\title{Combined two-loop self-energy corrections at finite and zero temperatures}% Force line breaks with \\

\author{ T. Zalialiutdinov $^{1,\,2}$}
\email[E-mail:]{t.zalialiutdinov@spbu.ru}
\author{D. Solovyev $^{1,\,2}$}
\affiliation{ 
$^1$ Department of Physics, St. Petersburg State University, Petrodvorets, Oulianovskaya 1, 198504, St. Petersburg, Russia\\
$^2$ Petersburg Nuclear Physics Institute named by B.P. Konstantinov of National Research Centre 'Kurchatov Institut', St. Petersburg, Gatchina 188300, Russia\\
}

\date{\today}

\begin{abstract}
In this paper we investigate higher-order corrections to the energies of bound states in hydrogen subjected to the external blackbody radiation field. In particular, within the framework of thermal quantum electrodynamics and $S$-matrix approach we analyze combined type of two-loop self-energy corrections, including one zero-vacuum and one loop at finite temperature. By utilizing the method of dimensional regularization, we derive closed analytical expressions for the energy shifts of atomic levels. Our numerical calculations demonstrate that even at room temperature these corrections can be significant for excited states, reaching the magnitude of the thermal induced Stark contribution.
\end{abstract}

\maketitle

\section{Introduction}

The interaction between blackbody radiation (BBR) and atomic systems has been a subject of interest %among physicists
for many years \cite{Gallagher:1979:PRL, Gallagher:1980:PRA, Beiting1979, Gallagher:1979:APL, Farley}. Recent advancements in atomic physics have revealed the significance of BBR-stimulated effects in both fundamental and applied sciences \cite{Brandt:2022:PRL}. In the pursuit of greater precision in the measurement of atomic transition energies and the development of frequency standards, the impact of uncertainty caused by BBR cannot be ignored, as evidenced by the most accurate clock experiments \cite{Saf1, Saf3, Kozlov-RevModPhys} and frequency measurements \cite{Nez, Schwob, Brandt:2022:PRL, deB-0, deB-1, deB-2}.  This challenge has prompted extensive research on the calculation of BBR-induced shift in clock systems \cite{Mid, Porsev}, with the potential to revolutionize the field of high-precision metrology \cite{riehle2006frequency}.

While early research on BBR-induced effects focused primarily on Rydberg atoms \cite{Hall}, the emergence of high-precision spectroscopy and frequency metrology has expanded the study of thermal induced effects to low-lying energy levels \cite{Itano, Brandt:2022:PRL, Jentschura:BBR:2008}. This development offers a promising path towards understanding fundamental physical constants, including the Rydberg constant, $R_{\infty}$, and fine-structure constant, $\alpha$. As the effects are not very pronounced, consideration of the finite temperature impact on atomic systems is typically limited to lower-order corrections within the framework of quantum mechanical (QM) perturbation theory. 

Previously, we developed a method for calculating higher-order corrections in the framework of the $S$-matrix line profile approach \cite{Andr, ZSLP-report} and quantum electrodynamics of bound states at finite temperatures (TQED), which makes it possible to carry out calculations in complete analogy with ordinary quantum electrodynamics at zero temperature. In particular, thermal one-loop corrections to hyperfine splitting, $g$-factor, recombination cross sections, and probabilities of one- and two-photon transitions were calculated \cite{SZA_2021, ZAS-2l, ZSL-1ph, SZATL, ZGS:gfactor:2022, ZGS:HFS:2022, SZATL}.

In this study, we extend the application of these methods to calculate combined two-loop self-energy radiative corrections with one ordinary and one thermal loop to the energy levels of a hydrogen-like atom. Our approach for evaluating the relevant equations incorporates the bound-state $S$-matrix formalism, finite-temperature quantum field theory, and non-relativistic quantum electrodynamics (NRQED), together with a technique known as dimensional regularization. The investigation of these higher-order corrections can be essential for achieving more accurate determinations of the blackbody radiation (BBR) shift in various physical systems. Our results pave the way for further progress in this field.

The paper is structured as follows. In Section \ref{section:1}, we consider the derivation of the leading order one-loop self-energy contribution to the energy shift by applying the dimensional regularization approach. Section \ref{section:2} is dedicated to the evaluation of the finite temperature one-loop contribution and its renormalization. In Section \ref{MAIN}, we apply these approaches to evaluate the combined two-loop problem. We present the numerical results and discussion in Section \ref{final}. Throughout the paper, we use relativistic units in which the reduced Planck constant $\hbar$, the speed of light $c$, and the vacuum permittivity $\varepsilon_{0}$ are set to unity ($\hbar=c=\varepsilon_{0}=1$). The fine structure constant $\alpha$ is given in these units as $\alpha = e^2/(4\pi)$, where $e$ is the electron charge.

\section{One-loop electron self-energy in the dimensional regularization}
\label{section:1}

In this section we briefly remind the derivation of leading order $\alpha (\alpha Z)^4$ ($Z$ is the nuclear charge) self-energy correction to the atomic energy level within the nonrelativistic approach applying dimensional regularization.  As is customary in dimensionally regularized quantum electrodynamics, we assume that the dimension of the space-time is $D = 4 - 2\varepsilon$, and that of space $d = 3 -2\varepsilon$. The parameter $\varepsilon$ is considered as small, but only on the level of matrix elements, where an analytic continuation to a noninteger spatial dimension is allowed. The regularization procedure used throughout this work relates to the modified minimal subtraction scheme, $\overline{\text{MS}}$, which absorbs the divergent part plus a universal constant that always arises along with the divergence in Feynman diagram calculations into the counterterms. When using dimensional regularization, i.e., $d^4 k \to \mu^{4-d} d^d k$, it is implemented by rescaling the renormalization scale: $\mu^2 \to \mu^2 \frac{e^{\gamma_{E}}}{4 \pi}$, with $\gamma_{E}$ being the Euler–Mascheroni constant.

\begin{figure}[hbtp]
    \centering
    \includegraphics{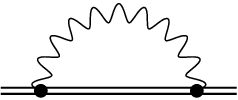}
    \caption{The Feynman graphs representing the lowest-order self-energy QED correction to the atomic energy level. The double solid line denotes the electron in the external Coulomb potential $V$ (the Furry picture), the wavy line denotes the virtual photon.}
    \label{fig:1}
\end{figure}

At first we briefly discuss the extension of the basic formulas of NRQED to the case of an arbitrary number of dimensions. The application of dimensional-regularized NRQED approach to estimate the Lamb shift in hydrogen can be found in \cite{PINEDA1998391}. The energy shift of the state $a$ corresponding to the one-loop Feynman diagram depicted in Fig.~\ref{fig:1} is given by the following expression \cite{Pachucki:1996}:
\begin{eqnarray}
\label{xa1}
\Delta E_{a} = -\mathrm{i}e^2\int%\limits_{C}
\frac{d^D K}{(2\pi)^D}D_{\mu\nu}(K)
\\\nonumber
\times
\langle \overline{\psi}_{a} |\gamma^{\mu}\frac{1}{\slashed{p}-\slashed{K}-m-\gamma_{0}V}\gamma^{\nu}|\psi_{a}\rangle -\delta m 
,
\end{eqnarray}
where 
\begin{eqnarray}
    \label{vacuumprop}
    D_{\mu\nu}(K)=\frac{g_{\mu\nu}}{K^2}
\end{eqnarray}
is the photon propagator in the Feynman gauge ($\mu,\,\nu=0,\,1,\,2,\,3$), $K=(k_{0},\bm{k})$ and $p=(p_{0},\bm{p})$ are the the 4-vectors of photon and electron momenta, respectively, $e$ is the electron charge, $g_{\mu\nu}$ is the metric tensor, $\delta m$ is the one-loop mass counter term, $\psi_{a}$ is the solution of Dirac equation for the hydrogen atom and $\overline{\psi}_{a}=\gamma_{0}\psi_{a}^{\dagger}$ its Dirac conjugation. The Coulomb potential $V$ in the denominator of Eq.~(\ref{xa1}) is given by 
\begin{eqnarray}
\label{coulomb1}
V(\bm{q})=-\frac{Ze^2}{q^2}
,
\end{eqnarray}
where $q=+\sqrt{\bm{q}^2}$. The Fourier transform of Eq.~(\ref{coulomb1}) can be written as follows
\begin{eqnarray}
\label{coulomb2}
V(\bm{r})=-Ze^2\int\frac{d^dq}{(2\pi)^d}\frac{e^{\mathrm{i}\bm{q}\bm{r}}}{q^2}
=
-\frac{Z_{\varepsilon}e^2}{4\pi r}
%\\\nonumber
=-\frac{Z_{\varepsilon}\alpha}{r}
.
\end{eqnarray}
The latter representation provides an implicit definition
of $Z_{\varepsilon}$ \cite{Jentschura_NRLAMB:2005}. The integration over $k_{0}$ in Eq. (\ref{xa1}) is taken along the standard Feynman contour $C$ \cite{Greiner}.

% \begin{figure}
%     \centering
%     \includegraphics[scale =0.6]{Fig0.eps}
%     \caption{Integration contour $C_{0}$ in $k_{0}$ plane. Arrows on the contour define the pole-bypass rule. The poles $\pm |\bm{k} |$ are denoted with X marks.}
%     \label{fig:1}
% \end{figure}

To pass to the  nonrelativistic limit, we first transform the matrix element in the integrand of Eq.~(\ref{xa1}) \cite{jentschura2003theory}:
\begin{align}
\label{xa2}
& 
\langle \overline{\psi}_{a} |\gamma^{\mu}\frac{1}{\slashed{p}-\slashed{K}-m-\gamma_{0}V}\gamma^{\nu}|\psi_{a}\rangle \qquad
\\\nonumber
% & 
% = 
% \langle \overline{\psi}_{a} |e^{\mathrm{i}\bm{k}\bm{r}}\gamma^{\mu}\frac{1}{\slashed{p}-\gamma_{0}k_{0}-\gamma_{0}V-m}\gamma^{\nu}e^{-\mathrm{i}\bm{k}\bm{r}}|\psi_{a}\rangle 
% \\\nonumber
& 
= 
\langle \overline{\psi}_{a} |\gamma^{\mu}e^{\mathrm{i}\bm{k}\bm{r}}\frac{1}{\slashed{p}-\gamma_{0}k_{0}-\gamma_{0}V-m}\gamma^{\nu}e^{-\mathrm{i}\bm{k}\bm{r}}|\psi_{a}\rangle
\\\nonumber
& 
= 
%\langle \psi_{a}^{\dagger} %|\gamma_{0}\gamma^{\mu}e^{\mathrm{i}\bm{k}\bm{r}}\frac{1}{(\gamma_{0}p_{0}-\gamma_{0}k_{0}-\bm{\gamma}\bm{%p}-\gamma_{0}V-m)}\gamma_{0}^{-1}\gamma_{0}\gamma^{\nu}e^{-\mathrm{i}\bm{k}\bm{r}}|\psi_{a}\rangle
%\\\nonumber
%& \qquad
%= 
\langle \psi_{a}^{\dagger} |\alpha^{\mu}e^{\mathrm{i}\bm{k}\bm{r}}\frac{1}{p_{0}-k_{0}-\bm{\alpha}\bm{p}-V-\gamma_{0}m}\alpha^{\nu}e^{-\mathrm{i}\bm{k}\bm{r}}|\psi_{a}\rangle
\\\nonumber
& 
= 
\langle \psi_{a}^{\dagger} |\alpha^{\mu}e^{\mathrm{i}\bm{k}\bm{r}}\frac{1}{E_{a}-k_{0}-H_{D}}\alpha^{\nu}e^{-\mathrm{i}\bm{k}\bm{r}}|\psi_{a}\rangle
,
\end{align}
where in the last line we took into account that $p_{0}=E_{a}$ is the Dirac energy of the bound state $a$ and $H_{D}=\bm{\alpha}\bm{p}-V-\gamma_{0}m$ is the Dirac Hamiltonian. 

In the Coulomb gauge the photon propagator is 
\begin{eqnarray}
\label{xa4}
D_{00}=\frac{1}{\bm{k}^2},
\end{eqnarray}
\begin{eqnarray}
\label{xa5}
D_{ij}=\frac{1}{K^2}\left(\delta_{ij} - \frac{k_{i}k_{j}}{\bm{k}^2}\right)
\end{eqnarray}
The integration over $k_0$ in Eq.~(\ref{xa1}) along the contour $C$ can be reduced to two scales in the self-energy problem: the atomic energy scale $m(\alpha Z)^2$ (low-energy part, $\Delta E_{a}^{\mathrm{L}}$) and the relativistic electron mass scale (high-energy part, $\Delta E_{a}^{\mathrm{H}}$). %The dimensional regularization allows a natural separation of energy scales using only one regularization parameter:
Then the dimensional regularization can be applied to both using only one regularization parameter: the dimension of the coordinate space, given by $d=3-2\varepsilon$. This leads to a straightforward derivation of radiative corrections in terms of the expectation values of effective operators and the Bethe logarithm. Following the work \cite{NRtwoloopJENTS}, we represent total one-loop self-energy correction to the atomic energy of the state $a$ as a sum of two contributions $\Delta E_{a}=\Delta E_{a}^{\mathrm{L}} + \Delta E_{a}^{\mathrm{H}}$.

The leading nonrelativistic low-energy contribution of Eq.~(\ref{xa1}) comes from the dipole approximation of matrix element given by Eq.~(\ref{xa2}). Performing integration over $k_{0}$ in Eq.~(\ref{xa1}) and taking into account that poles of electron propagator do not contribute in the low-energy limit \cite{PINEDA1998391,NRtwoloopJENTS}, we find
\begin{eqnarray}
\label{xa6}
\Delta E_{a}^{\mathrm{L}} = e^2\int\frac{d^dk}{(2\pi)^d}\frac{1}{2k}
\left(\delta_{ij} - \frac{k_{i}k_{j}}{\bm{k}^2}\right)
\\\nonumber
\times
\langle \phi_{a} |p_{i}
\frac{1}
{
E_a-H_{S}-k
}p_{j}|\phi_{a}\rangle
 ,
\end{eqnarray}
where $k=+\sqrt{\bm{k}^2}\equiv \omega$ and $H_{S}$ denotes the nonrelativistic Hamiltonian in $d$ dimensions. The wave function $\phi$, in contrast to $\psi$ in Eq.~(\ref{xa2}), corresponds to the nonrelativistic Schr\"odinger wave function. 
% In the a basis-set representation the matrix element in Eq.~(\ref{a6}) is
% \begin{eqnarray}
% \label{a7}
% \langle \phi_{a} |p_{i}
% \frac{1}
% {
% E_{a}-H_{S}-k
% }p_{j}|\phi_{a}\rangle
% \\\nonumber
% =
% \sum\limits_n
% \frac{\langle \phi_{a} |p_{i}|\phi_n\rangle \langle \phi_n |p_{j}|\phi_{a}\rangle}
% {
% E_a-E_n-k
% }
% \end{eqnarray}

The angular integration in Eq.~(\ref{xa6}) can be performed by noting that \cite{Yelkhovsky2001}
\begin{eqnarray}
\label{dint}
\int d^dk \left(\delta_{ij} - \frac{k_{i}k_{j}}{\bm{k}^2}\right)=
\int d\Omega_{d}k^{d-1}dk \left(\delta_{ij} - \frac{k_{i}k_{j}}{\bm{k}^2}\right)
\\\nonumber
=
\frac{2\pi^{d/2}}{\Gamma(d/2)}\frac{d-1}{d}
\delta_{ij}
\int\limits_{0}^{\infty} k^{d-1}dk
.
\end{eqnarray}
Then, using Eq.~(\ref{dint}) and expanding the result into the Taylor series up to the terms $\sim O(\varepsilon)$ (details are given in Appendix \ref{appendix:A}), we obtain
\begin{align}
\label{a8}
& 
\Delta E_{a}^{\mathrm{L}} = 
\frac{2\alpha}{3\pi}
\langle \phi_{a} |p_{i}
(H_{S}-E_{a})
\left[\frac{1}{2\varepsilon}
+\frac{5}{6}
-\frac{\gamma_{E}}{2}
\right.
\\
\notag
& 
\left.
-\log [2(H_{S}-E_{a})]+\frac{1}{2}\log 4\pi
\right]
p_{i} | \phi_{a} \rangle
.
\end{align}

The expression (\ref{a8}) can be simplified by noting that 
\begin{align}
\label{xa9}
& 
\langle 
\phi_{a} |
p_{i}
(H_{S}-E_{a})
p_{i}
| \phi_{a} 
\rangle
\\\nonumber
& 
=
\frac{1}{2}
\langle 
\phi_{a} |
[p_{i},[
H_{S},
p_{i}]]
+p^2H_{S} + H_{S}p^2
| \phi_{a} 
\rangle
\\\nonumber
& 
-E_{a}\langle 
\phi_{a} |p^2 |\phi_{a}\rangle
=
\frac{1}{2}
\langle 
\phi_{a} |
[p_{i},[
H_{S},
p_{i}]]
|\phi_{a}\rangle
\\\nonumber
& 
=
\frac{1}{2}
\langle 
\phi_{a} |
\Delta V
| \phi_{a} 
\rangle 
.
\end{align}
Then it is easy to show that the term in Eq. (\ref{xa9}), which is singular in the limit $\varepsilon\rightarrow 0$, is $+\frac{\alpha}{6\pi\varepsilon}\langle \phi_{a}|\Delta V|\phi_{a}\rangle$. Below we will see that the same divergence, but with the opposite sign, occurs in the high-energy part and is eventually compensated in the total shift. 

Taking into account that potential $V$ satisfies the $d$-dimensional Poisson equation
\begin{eqnarray}
\Delta V(\bm{r}) = 4\pi Z \alpha \delta^{(d)}(\bm{r}),
\end{eqnarray}
we arrive at
\begin{eqnarray}
\label{xa10}
\Delta E_{a}^{\mathrm{L}} = 
\frac{2\alpha}{3\pi}
\left[
 \langle \phi_{a} |2\pi Z\alpha\delta^d(\bm{r}) |\phi_{a}\rangle
\left(
\frac{5}{6}
-
\log (\alpha Z)^2
\right)
\right.
\\\nonumber
\left.
- \langle \phi_{a} |
p_{i}(H_{S}-E_{a})\log \frac{2(H_{S}-E_{a})}{(\alpha Z)^2}p_{i}
| \phi_{a} \rangle
\right].
\end{eqnarray}
Here the constants $-\frac{1}{2}\log 4\pi$ and $-\frac{\gamma_{\mathrm{E}}}{2}$ are removed by corresponding contribution in mass counter term $\delta m$ \cite{lindgren}. The matrix elements of operators arising in Eq.~(\ref{xa10}) are
\begin{eqnarray}
\label{xa11}
 \langle \phi_{a} |2\pi Z\alpha\delta^d(\bm{r}) |\phi_{a}\rangle=\frac{2(Z\alpha)^4}{n_{a}^3}\delta_{l0},
\end{eqnarray}
\begin{eqnarray}
\label{xa12}
\langle \phi_{a} |
p_{i}(H_{S}-E_{a})\log \frac{2(H_{S}-E_{a})}{(\alpha Z)^2}p_{i}
| \phi_{a} \rangle
\\\nonumber
= \frac{2 (\alpha Z)^4}{n_{a}^3}
\log \beta_{a}
,
\end{eqnarray}
where $\log \beta_{a}$ is the Bethe logarithm. The latter can be conveniently calculated in the acceleration gauge, see \cite{Tang:bethe:2013, Goldman:bethe:2000}, as follows:
\begin{eqnarray}
\label{bethe1}
\log \beta_{a} = \frac{B_{a}}{C_{a}}
,
\end{eqnarray}
where
\begin{eqnarray}
\label{bethe2}
B_{a}=\sum\limits_{n'l'}\frac{\left|\langle n_{a}l_{a} | \frac{\bm{r}}{r^3} |n'l'\rangle \right|^2 \log\frac{2|E_{n'}-E_{a}|}{(\alpha Z)^2}}
{E_{n'}-E_{a}}
\end{eqnarray}
\begin{eqnarray}
\label{bethe3}
C_{a}=\sum\limits_{n'l'}\frac{\left|\langle n_{a}l_{a} | \frac{\bm{r}}{r^3} |n'l'\rangle\right|^2 } {E_{n'}-E_{a}}
.
\end{eqnarray}
% \vbox{
% \begin{align}
% \label{a12}
% & \qquad
% \log \beta_{a} = \frac{n_{a}^3}{2(\alpha Z)^4}
% \left\lbrace
% \frac{l_{a}}{l_{a}+1}\sum\limits_{n'}(E_{n'}-E_{a})^3 \right.
% \\
% \nonumber
% & \qquad
% \times
% \log\frac{2|E_{n'}-E_{a}|}{(\alpha Z)^2}
% \left(
% \int\limits_{0}^{\infty}dr r^3 R_{n_al_a}(r)R_{n'l_{a}-1}(r)
% \right)^2
% \\
% \nonumber
% & \qquad
% + \frac{l_{a}+1}{2l_{a}+1}\sum\limits_{n'}(E_{n'}-E_{a})^3
% \\
% \nonumber
% & \qquad
% \times
% \left.
% \log\frac{2|E_{n'}-E_{a}|}{(Z\alpha)^2}
% \left(
% \int\limits_{0}^{\infty}dr r^3 R_{n_al_a}(r)R_{n'l_{a}+1}(r)
% \right)^2
% \right\rbrace
% \end{align}
% }
Here the summation runs over the entire spectrum including discrete and continuum states. Equation (\ref{bethe1}) has a numerical advantage for the evaluation of Bethe logarithm since it accelerates the convergence of sums in Eqs. (\ref{bethe2}), (\ref{bethe3}). Using the B-spline approach \cite{DKB} for the solution of Schr\"odinger equation for hydrogen atom the infinite sums over entire spectrum can be converted to a finite sums over the pseudo states. Our numerical calculations of Bethe logarithm for $1s$ and $2s$ states in hydrogen leads to the values $2.984129$ and $2.811770$, respectively, which are consistent with \cite{Drake:BetheLog:1990}. 

Finally, substituting Eqs. (\ref{xa11}) and (\ref{xa12}) into Eq.~(\ref{xa10}) we arrive at
\begin{eqnarray}
\label{low}
\Delta E_{a}^{\mathrm{L}} = 
\frac{2\alpha}{3\pi}
\left[
\frac{2(Z\alpha)^4}{n_{a}^3}\delta_{l0}
\left(
\frac{5}{6}
-
\log (\alpha Z)^2
\right)
\right.
\\\nonumber
\left.
- \frac{2 (\alpha Z)^4}{n_{a}^3}
\log \beta_{a}
\right]
,
\end{eqnarray}

The next step in the consideration of one-loop self energy correction is the evaluation of high-energy part of Eq.~(\ref{xa1}). For the photon energies of the order of electron rest mass the electron propagator in Eq.~(\ref{xa1}) can be expanded in a series in powers of interaction with the Coulomb field of nucleus:
\begin{eqnarray}
\label{series}
\frac{1}{\slashed{p}-\slashed{K}-m-\gamma_{0}V} = \frac{1}{\slashed{p}-\slashed{K}-m}
\\\nonumber
+
\frac{1}{\slashed{p}-\slashed{K}-m}\gamma_{0}V
\frac{1}{\slashed{p}-\slashed{K}-m}
+\dots
\end{eqnarray}

\begin{figure}[hbtp]
    \centering
    \includegraphics[scale = 1.7]{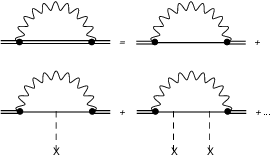}
    \caption{Potential expansion for the electron self-energy radiative correction. The ordinary solid line corresponds to the free electron, the dashed line with the cross at the end denotes the external potential.}
    \label{fig:3}
\end{figure}
Substituting Eq.~(\ref{series}) into Eq.~(\ref{xa1}) the high-energy part in the leading order can be written as a sum of zero and one potential terms (the first and second graphs in the right-hand side of the equation in Fig.~\ref{fig:3}):
\begin{eqnarray}
 \Delta E_{a}^{\mathrm{H}} = \langle \overline{\psi}_{a}| \Sigma(p) |\psi_{a} \rangle  + \langle \overline{\psi}_{a}| \Gamma_{0}(p,p)V|\psi_{a} \rangle -\delta m
 ,
\end{eqnarray}
where $\Gamma_{0}(p',p)$ is the vertex-function \cite{Greiner}
\begin{eqnarray}
\label{b03}
\Gamma_{\delta}(p',p)=-\mathrm{i}e^2\int%\limits_{C}
\frac{d^DK}{(2\pi)^D}\frac{g_{\mu\nu}}{K^2}
\\\nonumber
\times
\left(
\gamma^{\mu}  
\frac{1}{\slashed{p}\,'-\slashed{K}-m}
\gamma_{\delta}
\frac{1}{\slashed{p}-\slashed{K}-m}
\gamma^{\nu}
\right)
,
\end{eqnarray}
and $\Sigma(p)$ is the free-electron self-energy
\begin{eqnarray}
\Sigma(p)=-\mathrm{i}e^2 \int%\limits_{C} 
\frac{d^DK}{(2\pi)^D }\frac{g_{\mu\nu}}{K^2}
\gamma^{\mu}
\frac{1}{\slashed{p}-\slashed{K}-m}
\gamma^{\nu}
.
\end{eqnarray}
Note that higher order terms of decomposition (\ref{series}) do not contribute at $\alpha (\alpha Z)^4$ level.

Now we decompose $\Gamma_{\mu}(p',p)$ into a sum of the limit for zero momentum transfer $q=p'-p=0$ ("forward scattering") and the remainder \cite{Greiner}:
\begin{eqnarray}
\label{b04}
\Gamma_{\mu}(p',p) = \Gamma_{\mu}(p,p) + \Gamma^{\mathrm{R}}_{\mu}(p',p).
\end{eqnarray}
% Using the Ward identity the first term $\Gamma_{\mu}(p,p) $ in Eq.~(\ref{b4}) can be expressed as follows
% \begin{eqnarray}
% \label{b05}
% \Gamma_{\mu}(p,p)= - \frac{\partial}{\partial p^{\mu}}\Sigma (p)
% .
% \end{eqnarray}
The dimensionally regularized vertex-function $\Gamma^{\mathrm{R}}_{\mu}(p',p)$  is given by the electron form factors $F_1(q^2 )$ and $ F_2 (q^2) $:
\begin{eqnarray}
\label{gamR}
\Gamma^{\mathrm{R}}_{\mu}(p',p) = F_1 (q^2) \gamma_{\mu}  +
 \frac{\mathrm{i}}{2}  F_2 (q^2) \sigma_{\mu\nu} q^{\nu},
\end{eqnarray}
where 
\begin{align}
\label{b07}
&
F_1(q^2)=\frac{\alpha}{2\pi}\left[-\frac{1}{3\varepsilon}
-\frac{1}{4} - \varepsilon \right]q^2 +O(q^4)
,
\end{align}
\begin{align}
\label{b07-q}
&
F_2(q^2)=\frac{\alpha}{2\pi}
\left[
\left(1+4\varepsilon \right) 
+ 
\left(
\frac{1}{6}
+\frac{5}{6}\varepsilon
\right)
q^2
\right]
+O(q^4)
,
\end{align}
and
\begin{eqnarray}
\label{b06}
\sigma_{\mu\nu} = \frac{\mathrm{i}}{2}[\gamma_{\mu},\gamma_{\nu}]
.
\end{eqnarray}
After the cancellation of singular part of vertex-function and free-electron self-energy with mass counter term $\delta m$ the leading order of high-energy part is
\begin{eqnarray}
\label{a2h}
\Delta E_{a}^{\mathrm{H}}  =
\langle \overline{\psi}_{a} |
\Gamma^{\mathrm{R}}_{0}(p,p)V
|\psi_{a}\rangle 
\\\nonumber
=
\langle \overline{\psi}_{a} |
F_1 (q^2) \gamma_{0}V  +
 \frac{\mathrm{i}}{2}  F_2 (q^2) \sigma_{0\nu} q^{\nu}V
|\psi_{a}\rangle 
,
\end{eqnarray}
where the regularized vertex-function $\Gamma_{0}^{\mathrm{R}}$ contains only infrared divergence in parameter $\varepsilon$. For further evaluation it is convenient to rewrite Eq.~(\ref{a2h}) in the coordinate representation:
\begin{align}
\label{h3}
&
 \Delta E_{a}^{\mathrm{H}}  =
\langle \overline{\psi}_{a} |
\frac{\alpha}{2\pi}
\left(-\frac{1}{3\varepsilon} - \frac{1}{4}
\right) \gamma_{0}\Delta V  
\\\nonumber
&
+
 \frac{\mathrm{i\alpha}}{4\pi} (\bm{\alpha} \nabla V)
|\psi_{a}\rangle 
.
\end{align}
Here we used the fact that potential $V$ does not depend on time and substituted $q^2\rightarrow -(\partial_{t}^2-\Delta)\rightarrow \Delta $ together with the anti-commutation relation $\lbrace \gamma_{\mu},\gamma_{\nu} \rbrace=2g_{\mu\nu}$. Passing to the nonrelativistic limit in Eq.~(\ref{h3}) and using the Foldy–Wouthuysen transformation \cite{LabKlim}, we find
\begin{align}
\label{qwe}
&
\Delta E_{a}^{\mathrm{H}}  =\langle \phi_{a}|
\left[
\frac{\alpha}{2\pi}
\left(-\frac{1}{3\varepsilon} - \frac{1}{4}
\right)
\Delta V
\right.
\\\nonumber
&
\left. 
+\frac{\alpha}{8\pi}\Delta V
-
\frac{\alpha}{2\pi}\frac{1}{r}\frac{dV}{dr}(\bm{l}\cdot \bm{s})
\right]
|\phi_{a}\rangle
\\\nonumber
&
=
\langle \phi_{a}|
\left[-
\frac{\alpha}{6\pi\varepsilon}\Delta V 
% \right.
% \\\nonumber\times
% \left.
-
\frac{\alpha}{2\pi}\frac{1}{r}\frac{dV}{dr}(\bm{l}\cdot \bm{s})
\right]
|\phi_{a}\rangle
.
\end{align}
Note, that the divergent term in Eq.~(\ref{qwe}) is $-\frac{\alpha}{6\pi\varepsilon}\langle \phi_{a}|\Delta V|\phi_{a}\rangle$ and exactly cancels the same contribution in the low energy part, see Eq.~(\ref{a8}). 

For our further purposes it is also convenient to give an expression for the remaining regular part of matrix element in Eq.~(\ref{qwe}) which only contributes to the high-energy part in the nonrelativistic limit:
\begin{align}
\label{GVreg}
&
\langle \phi_{n'} |
\Gamma^{\mathrm{R}}_{0}(p,p)V
|\phi_{n}\rangle_{\mathrm{reg}} = -
\frac{\alpha}{2\pi}
\langle \phi_{n'} |
\frac{1}{r}\frac{dV}{dr}(\bm{l}\cdot \bm{s})
|\phi_{n}\rangle
\\\nonumber
&
=
-\frac{\alpha (Z\alpha)}{2\pi}
\langle \phi_{n'} |
r^{-3}(\bm{l}\cdot \bm{s})
|\phi_{n}\rangle
.
\end{align}
Performing the angular reduction of matrix element we have for $l_{a}\neq 0$:
\begin{eqnarray}
\label{high}
\Delta E_{a}^{\mathrm{H}}  = -\frac{\alpha(Z\alpha)^4 }{2\pi n_{a}^3}\frac{\left(j_{a} (j_{a}+1)-l_{a}(l_{a}+1)-\frac{3}{4}\right)}{l_{a}(l_{a}+1)(2l_{a}+1)}
\end{eqnarray}
Assembling the low-energy part (\ref{low}) and the high-energy part (\ref{high}), the finite result of the lowest order for one-loop self-energy is
\begin{align}
&
\Delta E_{a}^{\mathrm{SE}}  =\Delta E_{a}^{\mathrm{L}}   + \Delta E_{a}^{\mathrm{H}}  
\\\nonumber
&
=
\frac{4\alpha (Z \alpha)^4}{3\pi n_{a}^3}
\left[
\left(
\frac{5}{6}
-
2\log (\alpha Z)
\right)
\delta_{l_{a}0}
- 
\log \beta_{a}
\right]
\\\nonumber
&
-\frac{\alpha }{2\pi}  \frac{(Z\alpha)^4}{n_{a}^3} \frac{\left(j_{a} (j_{a}+1)-l_{a}(l_{a}+1)-3/4\right)}{l_{a}(l_{a}+1)(2l_{a}+1)}
.
\end{align}
This expression coincides with the well-known expression for the leading order of one-loop Lamb shift derived within Pauli-Villars regularization scheme \cite{Berest}. For the $1s$ and $2s$ states we find $8344.5$ MHz and $1066.4$ MHz, respectively.

\section{Thermal one-loop self-energy}
\label{section:2}

\begin{figure}[hbtp]
    \centering
    \includegraphics[scale = 0.6]{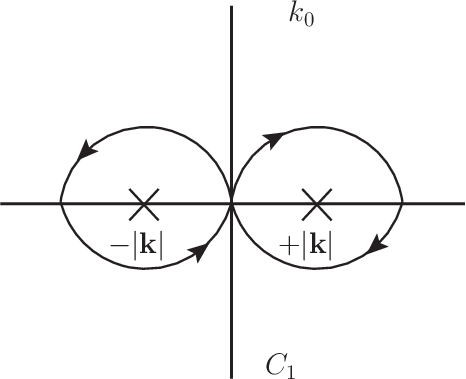}
    \caption{Integration contour $C_{1}$ in $k_{0}$ plane. Arrows on the contour define the pole-bypass rule. The poles $\pm |\bm{k} |$ are indicated by oblique crosses.}
    \label{fig_C1}
\end{figure}

In this section we briefly remind the derivation of thermal one-loop self-energy correction to the arbitrary atomic energy level $a$. In the leading order it is given by the well-known AC-Stark shift induced by the equilibrium radiation field with Planckian spectrum. The corresponding correction can be obtained by replacing ordinary photon propagator $D_{\mu\nu}$ in Eq.~(\ref{xa1}) by thermal one $D_{\mu\nu}^{\beta}$ \cite{DHR}.  In \cite{S-2020} it was found that the thermal part of photon propagator $D_{\mu\nu}^{\beta}$ \cite{DHR} admits
a different (equivalent) form which implies the integration over $k_0$ along the contour $C_{1}$, see Fig.~\ref{fig_C1}. 

In the space representation the finite temperature part of photon propagator reads
\begin{eqnarray}
\label{thermal_prop}
D_{\mu\nu}^{\beta}(x,x')= g_{\mu\nu}\int\limits_{C_{1}}\frac{d^4K}{(2\pi)^4} \frac{e^{\mathrm{i}K(x-x')}}{K^2}n_{\beta}(E_{k})
.
\end{eqnarray}
Here $n_{\beta}(E_{k}) =(\exp(E_{k}\beta)-1)^{-1} $ is the photon occupation number (Planck's distribution function), $E_{k}=|\bm{k}|$, $\beta =1/(k_{B}T)$, $k_{B}$ is the Boltzmann constant and $T$ is the radiation temperature in Kelvin. The equivalence of both forms, given by Eq. (\ref{thermal_prop}) and in \cite{DHR}, was demonstrated in \cite{S-2020}. Then the correction corresponding to the graph in Fig.~\ref{fig:1} can be expressed as
\begin{eqnarray}
\label{t3}
\Delta E_{a} = -\mathrm{i}e^2\int\limits_{C_{1}}\frac{d^4 K}{(2\pi)^4}D^{\beta}_{\mu\nu}(K)
\\\nonumber
\times
\langle \overline{\psi}_{a} |\gamma^{\mu}\frac{1}{\slashed{p}-\slashed{K}-m-\gamma_{0}V}\gamma^{\nu}|\psi_{a}\rangle -\delta m^{\beta}
,
\end{eqnarray}
where $\delta m^{\beta}$ is the thermal mass counter term.  As in the case of ordinary QED at $T=0$, it can be obtained through the diagram of the free thermal self-energy on the mass shell. An accurate evaluation of the thermal mass counter term was performed in \cite{DHR}, resulting in $\delta m^{\beta} = \pi \alpha (k_{B}T)^2/3$ (see Eq.~(35) there).% in \cite{DHR}). 

% It is important to note that since the vacuum expectation value of the particle number operator in thermal QED is nonzero, there is also an additional contribution to the thermal mass $\delta m^{\beta}$ due to the tadpole diagrams, see Fig. (\ref{}). Its contribution can be easily evaluated (see \cite{Escobedo2008, tadpole} ) and results to the additional term which is also $\frac{\pi \alpha}{3}(k_{B}T)^2$, therefore the total  thermal mass is $\delta m^{\beta}=\frac{2\pi \alpha}{3}(k_{B}T)^2$.

It should be noted that in contrast to the ordinary one-loop correction Eq.~(\ref{xa1}) the ultraviolet divergence is absent due to the factor $n_{\beta}$ in the integrand of Eq. (\ref{thermal_prop}). Hence, the contribution of the high-energy part, when $k_0$ is of the order of the electron rest-mass, is strongly suppressed and only the low-energy contribution can be considered. This also implies that the temperatures are sufficiently low, i.e $k_{B}T$ is of the order of binding energy $m(\alpha Z)^2$ or less. 

As for an ordinary self-energy loop, we pass to the Coulomb gauge and nonrelativistic limit for the operators and wave functions \cite{Jentschura_NRLAMB:2005}. Then integration over $k_0$ along the contour $C_{1}$ leads to  
\begin{eqnarray}
\label{beta6}
\Delta E_{a}^{\beta} = e^2\sum\limits_{\pm}\int\frac{d^3k}{(2\pi)^3}\frac{n_{\beta}(k)}{2k}
\left(\delta_{ij} - \frac{k_{i}k_{j}}{\bm{k}^2}\right)
\\\nonumber
\times
\langle \phi_{a} |p_{i}
\frac{1}
{
E_a-H_{S}\pm k
}p_{j}|\phi_{a}\rangle
-\delta m^{\beta}
 .
\end{eqnarray}
Here $\sum_{\pm}$ denotes the sum of two contributions with $+k$ and $-k$ in the denominator. Performing in Eq.~(\ref{beta6}) integration over the angles, Eq.~(\ref{dint}), we arrive at
\begin{eqnarray}
\label{xbeta7}
\Delta E_{a}^{\beta} = \frac{2\alpha}{3\pi}\sum\limits_{\pm}\int\limits_{0}^{\infty} dk\, k\, n_{\beta}(k)
\\\nonumber
\times
\langle \phi_{a} |p_{i}
\frac{1}
{
E_a-H_{S}\pm k
}p_{i}|\phi_{a}\rangle
-\delta m^{\beta}
 .
\end{eqnarray}
For numerical calculations it is convenient to rewrite Eq.~(\ref{xbeta7}) in the basis set representation and with matrix elements in the length form. The former can be done with the use of equality
\begin{eqnarray}
\label{bs}
\frac{1}
{
E_{a}-H_{S}-k
}
=
\sum\limits_n
\frac{|\phi_n\rangle \langle \phi_n |}
{
E_a-E_n-k}
,
\end{eqnarray}
where the sum over $n$ runs over entire spectrum of Schr\"odinger equation including continuum. The transition to the length form is made using the expression (\ref{aa2}) given in Appendix \ref{appendix:B}. The result is
\begin{align}
\label{beta7}
& \qquad
\Delta E_{a}^{\beta} = \frac{2\alpha}{3\pi}\sum\limits_{\pm}\sum\limits_{n}\int\limits_{0}^{\infty}  dk\, k^3\, n_{\beta}(k)
\\\nonumber
& \qquad
\times
\frac{\langle \phi_{a} |r_{i}| \phi_{n}\rangle \langle \phi_{n} |r_{i}|\phi_{a}\rangle}
{
E_a-E_{n}\pm k
}
+\frac{2\alpha}{\pi}\int\limits_{0}^{\infty}  dk\, k\, n_{\beta}(k)
-\delta m^{\beta}
 .
\end{align}
The first term in Eq.~(\ref{beta7}) represents ordinary AC-Stark shift induced by the blackbody radiation field \cite{Farley}. The second term is state independent and can be evaluated analytically, leading to
\begin{eqnarray}
\label{beta8}
\frac{2\alpha}{\pi}\int\limits_{0}^{\infty}  dk\, k\, n_{\beta}(E_{k}) = \frac{\pi\alpha}{3}(k_{B}T)^2
.
\end{eqnarray}
% To this term we must also add the corresponding contribution from the tadpole diagram, but for bound electrons. Its calculation is completely similar to the calculation for free ones and just doubles the state independent term in Eq.~(\ref{beta7}). We also note that in ordinary QED at $T=0$ the tadpoles contributions are absent.  This can be easily demonstrated with the use of technique of dimensional regularization, see \cite{}. 

After substituting Eq.~(\ref{beta8}) into Eq.~(\ref{beta7}), the state-independent contribution is cancelled by the mass-counter term, $\delta m^{\beta}$. % and only the pure AC-Stark contribution remains:
Then the AC-Stark shift is
\begin{eqnarray}
\label{beta9}
\Delta E_{a}^{\beta} = \frac{2\alpha}{3\pi}\sum\limits_{\pm}\sum\limits_{n}\int\limits_{0}^{\infty}  dk\, k^3\, n_{\beta}(E_{k})
\\\nonumber
\times
\frac{\langle \phi_{a} |r_{i}| \phi_{n}\rangle \langle \phi_{n} |r_{i}|\phi_{a}\rangle}
{
E_a-E_{n}\pm k
}
.
\end{eqnarray}
This equation coincides with a well-known quantum mechanical result \cite{Farley}. Taking into account that in relativistic units $k\sim m(\alpha Z)^2$, $r\sim (m\alpha Z)^{-1}$ and $\int_{0}^{\infty}k^3n_{\beta}(k) \sim (k_{B}T)^4$ we find that for the ground state $\Delta E_{a}^{\beta} \sim \frac{(k_{B}T)^4}{m^3\alpha^3 Z^4}$ r.u., which is in agreement with estimations given in \cite{Jentschura:BBR:2008}. %At the same time, for states with $n\ge 2$, energy differences can arise in denominator of Eq. (\ref{beta9}) which are shifted relative to each other only by the Lamb shift
At the same time, for states with $n\ge 2$ in the denominator of Eq. (\ref{beta9}), energy differences corresponding to the Lamb shift can arise, see \cite{Jentschura:BBR:2008,SLP-QED}. In this case, the temperature parameter comes at a lower power, which increases the magnitude of the correction. At $T=300$ K the  AC-Stark shifts are $ -0.039 $ Hz and $ -1.04 $ Hz for the $1s$ and $2s$ states in the hydrogen atom, respectively. As the temperature increases to $T=1000$ K, the corresponding shifts become $-4.79$ Hz and $-132.2$ Hz. The accurate numerical evaluation of Eq. (\ref{beta9}) and the corresponding analysis can be found in \cite{Farley, SLP-QED, Jentschura:BBR:2008, Zalialiutdinov2022:JETP}. 

It should be noted that in higher orders $\delta m^{\beta}$, the thermal mass counter term has not yet been calculated, and the explicit separation of thermal mass in two-loop diagrams with a bound electron in the same manner is rather difficult. However, in the higher orders of perturbation theory, when passing from the velocity form to the length form, all terms proportional to $(k_{B}T)^2$ vanish in the sum of contributions from all diagrams depicted in Fig.~\ref{fig:5}. In particular, previous studies \cite{ZGS:gfactor:2022, ZGS:HFS:2022} have demonstrated the cancellation of such terms when calculating the thermal loop corrections to the bound electron $g$-factor or hyperfine splitting. %Thus, the efficient procedure for renormalizing of thermal loop in the nonrelativistic limit can be reduced to a calculation in the length form
Thus, the renormalization procedure of the thermal loop in the nonrelativistic limit can be validly reduced to a calculation in the length form, while the procedure of renormalization of zero-temperature (ordinary) loop remains the same.

%Therefore, for two-loop contributions, we will only be interested in the correction to the transition energy (i.e. energy difference) and use the form of expressions similar to the expression Eq.~(\ref{xbeta7_mod}).

% \begin{table}[hbtp]
%     \centering
%     \caption{Comparison of BBR-induced Stark shifts at $T=300$ K calculated with Eqs. (\ref{xbeta7}), (\ref{beta9}) and (\ref{xbeta7_mod}). All values are in Hz.}
%     \begin{tabular}{c c c c} 
%     \hline
%       State & Eq.~(\ref{xbeta7}) & Eq.~(\ref{beta9})  & Eq.~(\ref{xbeta7_mod})\\
%       \hline
%       $1s$   &  & 0.039 & 0.037\\
%       $2s$   & 1.017 & 1.043 & 1.033\\
%       \hline
%     \end{tabular}
%     \label{tab:1}
% \end{table}

\section{Two loop self-energy with one thermal loop}
\label{MAIN}

When dealing with two-loop combined-type diagrams, the calculation procedure can be separated into two contributions, similar to the one-loop case. It is important to note that in this scenario, the photon momenta in the thermal loop (denoted as $k_1$) are always in the low-energy region, since the temperatures under consideration are much lower than the electron rest mass. As a result, the high-energy contribution arises from the photon momenta of the ordinary loop (denoted as $k_2$). Consequently, the regularization process involves eliminating ultraviolet divergences in the integrals over the momentum $k_2$ of an ordinary photon. Meanwhile, the integral over the momenta of a thermal photon converges due to the presence of a Planckian distribution in the integrand.

Building upon the results of previous sections, we begin by considering the low-energy part of the two-loop problem. Specifically, we will extend the findings of the zero-temperature case discussed in \cite{NRtwoloopJENTS} to the case of finite temperature. To eliminate singular terms in this part the dimensional regularization is used. %On the other hand, in the high-energy regime, the same singular terms with opposite signs will emerge as the denominators, which contain the moment $k_2$, are expanded in powers of the Coulomb potential interaction in the limit of large photon momenta.
Following this technique, the same singular terms with opposite signs will arise in the high-energy regime when decomposing the denominators containing momentum $k_2$ by the powers of the Coulomb potential interaction in the limit of large photon momenta.

% The high-energy limit of the mixed two-loop correction to the energy shift can separated for two cases: 1) when energy of both photon loop momenta are of the order of electron rest mass, 2) when photon momenta of thermal loop is of the order $m(\alpha Z)^2$ while the momenta of ordinary loop is of the order of electron rest mass. The first case will be exponentially suppressed compared to the second one due to the presence of the Planck distribution function in the corresponding integrand.

% It is convenient to consider the high-energy contribution to the mixed two-loop problem in the Feynman gauge and to decompose propagators containing a non-thermal photon momenta in terms of the degrees of interaction with an external Coulomb field 

\subsection{Low-energy limit of two-loop contribution}

The low-energy contribution $\Delta E^{\mathrm{L}}_{a}$, which has been redefined for the two-loop problem, arises from two photon momenta that are of the order of $m\alpha^2$. Following the method described in \cite{NRtwoloopJENTS}, and taking into account Eq. (\ref{thermal_prop}) for the thermal photon propagator in the Coulomb gauge, $\Delta E_{a}$ the length gauge is expressed by
\begin{align}
\label{t1}
&
\Delta E_{a}^{\mathrm{L}}=
\left[e^2\int\limits \frac{ d^dk_1 k_1^2}{(2\pi)^d 2k_1}\frac{d-1}{d}n_{\beta}(k_1)\right] 
\\\nonumber
\times
&
\left[
e^2 \int\limits \frac{d^dk_2 k_2^2}{(2\pi)^d 2k_2}\frac{d-1}{d}
\right]
P(k_1,k_2),
\end{align}

\begin{widetext}
\begin{align}
\label{t2}
P(k_1,k_2)=\sum\limits_{\pm}
&
\left[
\left\langle 
\phi_{a}\left|r_i \frac{1}{E_{a}-H_{S}\pm k_1} r_j  \frac{1}{E_{a}-H_{S} \pm k_1+k_2} r_i  \frac{1}{E_{a}-H_{S}-k_2} r_j\right| \phi_{a}
\right\rangle_{\mathrm{CL}}
\right.
\\\nonumber
&
+
\left\langle
\phi_{a}\left|r_i \frac{1}{E_{a}-H_{S}\pm k_1} r_j  \frac{1}{E_{a}-H_{S} \pm k_1-k_2} r_j  \frac{1}{E_{a}-H_{S}\pm k_1} r_i\right| \phi_{a}
\right\rangle_{\mathrm{ViT}}
\\\nonumber
&
+
\left\langle
\phi_{a}\left|r_i \frac{1}{E_{a}-H_{S} - k_2} r_j  \frac{1}{E_{a}-H_{S} \pm k_1-k_2} r_j  \frac{1}{E_{a}-H_{S} - k_2} r_i\right| \phi_{a}
\right\rangle _{\mathrm{VoT}}
\\\nonumber
&
+ 
\left\langle 
\phi_{a}\left|r_i \frac{1}{E_{a}-H_{S}\pm k_1} r_i  \frac{1}{(E_{a}-H_{S})'} r_j  \frac{1}{E_{a}-H_{S} - k_2} r_j\right| \phi_{a}
\right\rangle _{\mathrm{LaL\,(irr)}}
\\\nonumber
&
-
\frac{1}{2}
\left\langle 
\phi_{a}\left|
r_i
\frac{1}{E_{a}-H_{S}\pm k_1}
r_i
\right| \phi_{a}
\right\rangle
\left\langle 
\phi_{a}\left|
r_j
\frac{1}{[E_{a}-H_{S}-k_2]^2}
r_j
\right| \phi_{a}
\right\rangle
 _{\mathrm{LaL\,(red\,1)}}
\\\nonumber
&
-
\frac{1}{2}
\left\langle 
\phi_{a}\left|
r_i
\frac{1}{E_{a}-H_{S}\pm k_2}
r_i
\right| \phi_{a}
\right\rangle
\left\langle 
\phi_{a}\left|
r_j
\left.
\frac{1}{[E_{a}-H_{S}-k_1]^2}
r_j
\right| \phi_{a}
\right\rangle _{\mathrm{LaL\,(red\,2)}}
\right] 
,
\end{align}
\end{widetext}
where $d=3-2\varepsilon$ and $d^dk= k^{d-1}dk d\Omega_{d}$. 

\begin{figure}[hbtp]
    \centering
    \includegraphics[scale = 1.9]{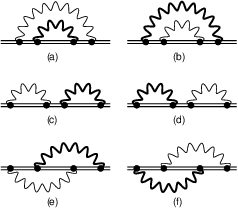}
    \caption{The Feynman graphs representing the combined two-loop self-energy QED corrections to the atomic energy level. Various contributions are indicated using the following notations: (a) - vacuum loop over thermal loop (VoT), (b) - vacuum loop inside thermal loop (ViT), (c) and (d) - loop-after-loop (LaL), (e) and (f) - crossed loops (CL). The double solid line denotes the electron in the external Coulomb potential $V$ (the Furry picture), the tiny wavy line denotes the zero-temperature virtual photon, while the bold one corresponds to the finite temperatures.}
    \label{fig:5}
\end{figure}

The terms in Eq.~(\ref{t2}) can be associated with different loop diagrams: the first term corresponds to the crossed loop diagrams (CL) depicted in Fig. \ref{fig:5} (e) and (f), the second term corresponds to the vacuum loop inside the thermal loop (ViT) diagram shown in Fig.~\ref{fig:5} (b); the third term represents the contribution of the vacuum loop over the thermal loop (VoT) diagram depicted in Fig.~\ref{fig:5} (a); while the last three summands corresponds to the irreducible and reducible parts of the loop after the loop (LaL), see Fig.~\ref{fig:5} (c) and (d). 

Taking into account that in relativistic units $e=\sqrt{4\pi\alpha}$, $k\sim m(\alpha Z)^2$, $dk\sim m(\alpha Z)^2$, $r\sim (m\alpha Z)^{-1}$ and $\int_{0}^{\infty}k^3n_{\beta}(k) = \frac{16\pi^5}{15}(k_{B}T)^4$ we find that for the ground state $\Delta E_{a}^{\beta} \sim \frac{(k_{B}T)^4}{m^3 Z^2}$ r.u. (or $\frac{16\pi^5}{15}\frac{\alpha^6(k_{B}T)^4}{Z^2}$ in a.u.). 
Neglecting the numerical coefficient arising in the matrix elements within the numerator, a parametric evaluation of the ground state of hydrogen ($Z = 1$) at a temperature $T = 300$ results in a preliminary estimate on the order of $10^{-7}$ Hz. Remarkably, this approximation exhibits significant congruence with results obtained through direct numerical computations, as will be subsequently demonstrated (see Table \ref{tab:1} below). At the same time, when integrating over $k_2$ or when summing over the spectrum, situations arise when one of the denominators reduces the power of $k_1$ in the numerator and, consequently, reduces power of the temperature in the corresponding parametric estimate thereby increasing the magnitude of the effect.

The integration over the $k_2$ can be done analytically using the dimensional regularization technique, while the integration over the $k_1$ will be done numerically in the last step of the calculation. All integrals necessary for these calculations are evaluated in the Appendix A. 

\subsubsection{Crossed loops}

In the Coulomb gauge and nonrelativistic limit the corresponding contribution is given by the first term in Eq. (\ref{t2}) and can be written as follows
\begin{align}
&
\Delta E^{\mathrm{L,\, CL}}_{a}=\sum\limits_{\pm}
\left[e^2\int\limits \frac{d^dk_1\,k_1^2}{(2\pi)^d 2k_1}\frac{d-1}{d}n_{\beta}(k_1)\right] 
\\\nonumber
&
\times
\left[
e^2 \int\limits \frac{d^dk_2\,k_2^2}{(2\pi)^d 2k_2}\frac{d-1}{d}
\right]
% \\\nonumber
% \times
\left\langle
\phi_{a}\left|r_i \frac{1}{E_{a}-H_{S} \pm k_1}
\right.
\right.
\\\nonumber
&
\times
\left.
\left.
r_j 
\frac{1}{E_{a}-H_{S} \pm k_1-k_2} r_i
\frac{1}{E_{a}-H_{S} - k_2} r_j\right| \phi_{a}
\right\rangle 
.
\end{align}
Before proceeding to isolating the finite part of the low-energy contribution, it is convenient to pass to the basis set representation, see equality (\ref{bs}). Then, performing $d$-dimensional integration over $k_2$ with the use of Eq.~(\ref{a3}), and angular integration over the $k_1$, the finite part of low-energy contribution of CL diagram is
\begin{align}
\label{finiteCL}
&
\Delta E^{\mathrm{L,\, CL}}_{a}=
\frac{2\alpha}{3\pi}    
\sum\limits_{\pm}
\sum\limits_{n_1n_2n_3}
\int\limits_{0}^{\infty}
dk_1 k_1^3n_{\beta}(k_1)
\\\nonumber
\times
&
\frac{
\langle
\phi_{a}|r_i |\phi_{n_1} \rangle 
\langle \phi_{n_1}
| 
r_j 
|\phi_{n_{2}} 
\rangle
\langle 
\phi_{n_{2}}| r_{i}|\phi_{n_{3}}\rangle 
\langle
\phi_{n_{3}}
|
r_j
| 
\phi_{a}
\rangle 
}{E_{a}-E_{n_1} \pm k_1}
\\\nonumber
&
\times
\frac{2\alpha}{3\pi}
\left\lbrace
\frac{5}{6}
\frac{
\left(
(E_{a}-E_{n_2}\pm k_1)^3-(E_{a}-E_{n_3})^3
\right)
}{E_{n_3}-E_{n_2}\pm k_1}
\right.
\\\nonumber
&
-\frac{(E_{a}-E_{n_2}\pm k_1)^3
\log[2|E_{a}-E_{n_2}\pm k_1|]}
{E_{n_3}-E_{n_2}\pm k_1}
\\\nonumber
&
\left.
+\frac{(E_{a}-E_{n_3})^3 \log[2|E_{a}-E_{n_3}|]}
{E_{n_3}-E_{n_2}\pm k_1}
\right\rbrace
,
\end{align}
where summation over $n_1,\,n_2,\,n_3$ extends over entire spectrum of Schr\"odinger equation for hydrogen atom including continuum states. 

The part with infrared divergence arising in the dimensional integration, see Eq.~(\ref{a3}), can be transformed with the use of commutation relation (\ref{HHR}) and resonance condition $k_{1}= \pm (E_{a}-E_{n_1})$ to the form:
\begin{eqnarray}
    \label{divCL0}
\frac{2\alpha}{3\pi}
\sum\limits_{\pm}\int\limits_{0}^{\infty} k_{1}^3 n_{\beta}(k_{1})
\left(\frac{\alpha}{6\pi\varepsilon} \right)
\\
\nonumber
\times
\sum\limits_{n_1}
\frac{\langle 
\phi_{a}
|
r_{i}  | \phi_{n_1} \rangle
\langle 
\phi_{n_1}
|
r_{i}
|
\phi_{n_2}
\rangle
\langle 
\phi_{n_2}
|
\Delta
|
\phi_{n_a}
\rangle
}
{E_{a}-E_{n_1}\pm k_1} 
. 
\end{eqnarray}
Below we show that it identically compensates for the similar contribution of the high-energy part, see Eq. (\ref{divCL}). Thus only finite result given by Eq.  (\ref{finiteCL}) remains.

\subsubsection{Vacuum loop inside thermal loop}

The corresponding contribution to the energy shift Eq.~(\ref{t1}) is given by the second term of Eq.~(\ref{t2}), which can be written as follows
\begin{eqnarray}
\Delta E^{\mathrm{L,\, ViT}}_{a}=\sum\limits_{\pm}
\left[e^2\int\limits \frac{d^dk_1\,k_1^2}{(2\pi)^d 2k_1}\frac{d-1}{d}n_{\beta}(k_1)\right] \times
\\\nonumber
\left[
e^2 \int\limits \frac{d^dk_2\,k_2^2}{(2\pi)^d 2k_2}\frac{d-1}{d}
\right]
% \\\nonumber
% \times
\left\langle
\phi_{a}\left|r_i \frac{1}{E_{a}-H_{S} \pm k_1}\times
\right.
\right.
\\\nonumber
%\times
\left.
\left.
r_j 
\frac{1}{E_{a}-H_{S} \pm k_1-k_2} r_j
\frac{1}{E_{a}-H_{S} \pm k_1} r_i\right| \phi_{a}
\right\rangle 
.
\end{eqnarray}
Integrating over $k_2$, see Eq.~(\ref{a1}), then over $k_1$ angles, and going to the basis set representation, we obtain
\begin{align}
\label{b009}
&
\Delta E^{\mathrm{L,\, ViT}}_{a}=
\frac{2\alpha}{3\pi}
\sum\limits_{\pm}
\sum\limits_{n_1n_2}
\int\limits_{0}^{\infty} dk_1 k_{1}^3n_{\beta}(k_1)\times
\\\nonumber
&
\frac{
\langle
\phi_{a}
|r_i| \phi_{n_1}\rangle 
\langle \phi_{n_1}| 
r_{j}
|
\phi_{n_2}\rangle 
\langle \phi_{n_2}| 
r_{j}
|
\phi_{n_3} \rangle 
 \langle \phi_{n_3} |r_i | \phi_{a}
\rangle }{(E_{a}-E_{n_1}\pm k_1)(E_{a}-E_{n_3}\pm k_1)}   \times
\\\nonumber
&
\left\lbrace
\frac{2\alpha}{3\pi}
(E_{n_2}-E_{a}\pm k_{1})^3
\left(
\frac{5}{6}
-
\log[2|E_{n_2}-E_{a}\pm k_1|]
\right)
\right\rbrace
.
\end{align}
%The singular term arising in the dimensional integral over $k_2$ (see Eq.~(\ref{a1})) is omitted in Eq.~(\ref{b009}). After conversion, Eq.~(\ref{off_xa9}) and applying the resonance condition $ k_1 = \pm(E_{a} - E_{n_1}) = \pm(E_{a} - E_{n_3})$ it can be written as

The singular term discarded above, after conversion Eq.~(\ref{off_xa9}) and application of the resonance condition $ k_1 = \pm(E_{a} - E_{n_1}) = \pm(E_{a} - E_{n_3})$, can be written as
\begin{align}
\label{ViTloweps}
&
\frac{2\alpha}{3\pi}\int\limits_{0}^{\infty} dk_{1}k_{1}^3n_{\beta}(k_{1})
\sum\limits_{\pm}
\sum\limits_{n_1n_2}
\left(+\frac{\alpha}{6\pi\varepsilon}\right)\times
\\\nonumber
&
\frac{
\langle 
\phi_{a}
|r_{i} |\phi_{n_{1}}\rangle \langle \phi_{n_{1}}|\Delta V|\phi_{n_{3}}\rangle \langle \phi_{n_{3}}|r_{i}| \phi_{a}
\rangle}{(E_{a}-E_{n_1}\pm k_{1})(E_{a}-E_{n_3}\pm k_{1})}.
\end{align}
Again, see below, the high-energy part of the ViT diagram represents the same contribution, but of the opposite sign. Thus, in the aggregate of the low-energy and high-energy parts, all infrared divergences arising in the renormalization procedure of this diagram disappear.

\subsubsection{Vacuum loop over thermal loop}

Within the nonrelativistic limit the contribution represented by the vacuum loop over the thermal loop, see the third term in Eq.~(\ref{t2}), is given by
\begin{align}
\label{VoTfinal}
&
\Delta E^{\mathrm{L,\, VoT}}_{a}=\sum\limits_{\pm}
\left[e^2\int\limits \frac{d^dk_1\,k_1^2}{(2\pi)^d 2k_1}\frac{d-1}{d}n_{\beta}(k_1)\right] 
\\\nonumber
&
\times
\left[
e^2 \int\limits \frac{d^dk_2\,k_2^2}{(2\pi)^d 2k_2}\frac{d-1}{d}
\right]
% \\\nonumber
% \times
\left\langle
\phi_{a}\left|r_i \frac{1}{E_{a}-H_{S} - k_2}
\right.
\right.
\\\nonumber
&
\times
\left.
\left.
r_j 
\frac{1}{E_{a}-H_{S} \pm k_1-k_2} r_j
\frac{1}{E_{a}-H_{S} - k_2} r_i\right| \phi_{a}
\right\rangle 
.
\end{align}
The integral over $k_2$ in Eq.~(\ref{VoTfinal}) diverges and within the dimensional regularization can be evaluated using Eq.~(\ref{a8}). Then the finite part is given by
\begin{widetext}   
\begin{eqnarray}
\Delta E^{\mathrm{L,\, VoT}}_{a}=
\frac{2\alpha}{3\pi}\int\limits_{0}^{\infty} dk_{1}k_{1}^3n_{\beta}(k_{1})
\left\lbrace
\frac{2\alpha}{3 \pi }
\langle
\phi_{a}
|r_i| \phi_{n_1}\rangle 
\langle \phi_{n_1}| 
r_{j}
|
\phi_{n_2}\rangle 
\langle \phi_{n_2}| 
r_{j}
|
\phi_{n_3} \rangle 
 \langle \phi_{n_3} |r_i | \phi_{a}
\rangle
\right.
\\\nonumber
\times
(3 E_{a}-E_{n_1}-E_{n_2}-E_{n_3}\pm k_{1})
\left(\frac{(E_{a}-E_{n_1})^3 \log [2
   |E_{a}-E_{n_1}|]-(E_{a}-E_{n_2}\pm k_{1})^3 \log [2
   |E_{a}-E_{n_2}\pm k_{1}|]}{(E_{n_1}-E_{n_2}\pm k_{1}) (-E_{n_2}+E_{n_3}\pm k_{1}) (3
   E_{a}-E_{n_1}-E_{n_2}-E_{n_3}\pm k_{1})}
   \right.
   \\\nonumber
   \left.
   \left.
   -\frac{(E_{a}-E_{n_1})^3 
   \log [2    |E_{a}-E_{n_1}|]-(E_{a}-E_{n_3})^3 \log [2 |E_{a}-E_{n_3}|]}{(E_{n_1}-E_{n_3})
   (-E_{n_2}+E_{n_3}\pm k_{1}) (3 E_{a}-E_{n_1}-E_{n_2}-E_{n_3}\pm k_{1})}+\frac{5}{6}
   \right)
   \right\rbrace
\end{eqnarray}
\end{widetext}

\subsubsection{Loop-after-loop}
\label{lal:low}

The last contribution to be considered is a loop-after-loop diagram. It can can be splitted into irreducible and reducible parts (reference state part). The latter can be represented as a product of the matrix elements of the diagonal one-loop matrix elements of the ordinary and thermal electron self-energy operators and the corresponding first derivatives over energy $E_{a}$ (last two terms in Eq. (\ref{t2})). Therefore, the evaluation of reducible loop-after-loop part is identical to the previously considered one-loop contribution and in this section we will focus on the low-energy limit of irreducible part only. This is
\begin{align}
\label{lal1}
&
\Delta E^{\mathrm{L,\, LaL}}_{a}(\mathrm{irr})=
\sum\limits_{\pm}
\left[
e^2\int\limits \frac{d^dk_1\,k_1^2}{(2\pi)^d 2k_1}\frac{d-1}{d}n_{\beta}(k_1)
\right]\times
\\\nonumber
&
\left[
e^2 \int\limits \frac{d^dk_2\,k_2^2}{(2\pi)^d 2k_2}\frac{d-1}{d}
\right]
\left\langle 
\phi_{a}
\left|
r_i \frac{1}{E_{a}-(H_{S}\pm k_1)} 
r_i \times
\right.
\right.
\\\nonumber
&
\left.
\left.
\frac{1}{(E_{a}-H_{S})^{'}}
r_j  
\frac{1}{E_{a}-(H_{S} + k_2)} 
r_j
\right| 
\phi_{a}
\right\rangle 
.
\end{align}
% The first term in curly brackets of Eq.~(\ref{lal1}) represents the irreducible contribution arising from the diagarn Fig. \ref{}, while the second corresponds to the reducible one (reference state term). Again performing $d$-dimensional integration over $k_2$, angular integration over the $k_1$ we arrive at

% \begin{widetext}
% \begin{eqnarray}
% \label{lal2}
% \Delta E^{\mathrm{L,\, LaL}}_{a}(\mathrm{irr})=
% \frac{2\alpha}{3 \pi }
% \sum\limits_{\pm}
% \int\limits dk_1 k_1^3 n_{\beta}(k_1)
% % \\\nonumber
% % \times
% \left\langle 
% \phi_{a}
% \left|
% r_i \frac{1}{E_{a}-(H_{S}\pm k_1)} r_j  
% \frac{1}{(E_{a}-H_{S})^{'}}
% r_i  
% \left\lbrace 
% \frac{2\alpha}{3 \pi } (H_{S}-E_{a})^3
% \right.
% \right.
% \right.
% \\\nonumber\times
% \left.
% \left(
% \frac{1}{2\varepsilon}-\log [2(H_{S}-E_{a})] 
% \right.
% \left.
% \left.
% \left.
% +\frac{5}{6} -\frac{\gamma_{\mathrm{E}}}{2} 
% +\frac{1}{2}\log 4\pi
% \right)
% \right\rbrace
% r_j
% \right| 
% \phi_{a}
% \right\rangle 
% .
% \end{eqnarray}
% \end{widetext}
Performing the same steps as in previous sections, the finite part in the basis set representation is reduced to
\begin{widetext}
\begin{eqnarray}
\label{lal3}
\Delta E^{\mathrm{L,\, LaL}}_{a}(\mathrm{irr})=
\frac{2\alpha}{3 \pi }
\sum\limits_{\pm}
\int\limits dk_1k_{1}^3n_{\beta}(k_1)
\left[
\frac{2\alpha}{3 \pi } 
\sum\limits_{\substack{n_1n_3\\n_2\neq a }}
\frac{
\langle 
\phi_{a}
|
r_i |  \phi_{n_1}\rangle \langle \phi_{n_1}|r_i |\phi_{n_2}\rangle
\langle \phi_{n_2} |r_j| \phi_{n_3}\rangle 
\langle \phi_{n_3} |
r_j
| 
\phi_{a}
\rangle 
}{(E_{a}-E_{n_1}\pm k_1)(E_{a}-E_{n_2})}  \times
\right.
\\\nonumber
\left.
\left(
\frac{5}{6}
(E_{n_3}-E_{a})^3
-(E_{n_3}-E_{a})^3\log [2|E_{n_3}-E_{a}|]
\right)
\right]
% \\\nonumber
% \left.
% +
% \frac{2\alpha}{3 \pi }
% \sum\limits_{n_1n_3}
% \frac{\langle 
% \phi_{a}|
% r_i|\phi_{n_1}\rangle \langle \phi_{n_1}|r_i
% | \phi_{a}
% \rangle
% \langle 
% \phi_{a}
% |r_j|
% \phi_{n_3}
% \rangle
% \langle
% \phi_{n_3}
% |r_j|
% \phi_{a}
% \rangle
% }{E_{a}-E_{n_1}\pm k_1}
% (E_{n_3}-E_{a})^2
% \log[2 |E_{n_{3}}-E_{a}|]
.
\end{eqnarray}
\end{widetext}
As before, the divergent contribution arising in the irreducible part is exactly compensated by a similar term in the irreducible high-energy part of the loop-after-loop diagram. 

\subsection{High-energy limit of two-loop contribution}
\label{HE}

In the two-loop combined diagrams we assume that the thermal photon energies are much smaller than the electron's rest mass, and the integrals over the closed thermal photon loop are suppressed by the Planck distribution. This means that the high-energy part is related only to the zero-temperature loop. Following the consideration of one-loop Lamb shift, we consider the two-loop high energy part in a similar way by denoting the finite and zero-temperature photon 4-momenta  as $K_1=(k_{10},\bm{k}_1)$ and $K_2=(k_{20},\bm{k}_2)$, respectively. 

To renormalize the thermal contribution in the nonrelativistic limit and the dipole approximation, we pass to the length form. After this step, we only need to renormalize the zero-temperature (ordinary) QED loop's contribution. We can achieve this by expanding all electron propagators that contain momenta $K_2$ in powers of the Coulomb interaction potential.

Similar to the one-loop case, we consider only the leading-order contributions in $\alpha Z$ and apply dimensional regularization. Doing this, we can extract an ultraviolet finite contribution after subtracting the mass counter-term that contains only the infrared divergence in the limit  $\varepsilon \rightarrow 0$. This divergence is compensated completely by the similar contribution of the low-energy part when the two energy scales are stitched together.

\subsubsection{Crossed loops}
\label{appB}

%The high energy part can be conveniently considered starting form the fully relativistic two-loop equations for energy shifts given in the Feynman gauge.
It is convenient to consider the high-energy part of the crossed loops relying on fully relativistic two-loop equations for energy shifts given in the Feynman gauge. Taking into account that ultraviolet divergences arise only when integrating over the photon momentum in a zero-temperature loop, the finite part can be obtained as in the one-loop case, i.e. expanding the denominators of the corresponding electron propagators by powers of the Coulomb interaction.

Then, performing $D=4-2\varepsilon$ dimensional integration over the photon loop and subtracting the corresponding mass counter-term, the final result can be expressed via the regularized vertex function $\Gamma_{\nu}^{\mathrm{R}}$ (see Eq. (\ref{gamR})) with the remaining regular temperature dependent part. %over the momenta of thermal photon loop. 
Thus, the infrared divergence is contained in $\Gamma_{\nu}^{\mathrm{R}}$. %, which is eliminated by stitching with the low energy part.

%We start the evaluation of high-energy limit with the crossed loops contribution which can be written as follows \cite{Sapirstein:2004, Yerokhin:PRL:2003}:
Following the above, see also \cite{Sapirstein:2004, Yerokhin:PRL:2003}, we write
\begin{widetext}
\begin{eqnarray}
\label{b1}
\Delta E^{\mathrm{H},\,\mathrm{CL}}_{a}=
-e^4\left[\int_{C_1} \frac{d^DK_1}{(2\pi)^D}\frac{g_{\mu\nu}}{K_{1}^2}n_{\beta}(E_{k_1})\right] 
\left[
    \int\frac{d^DK_2}{(2\pi)^D }\frac{g_{\lambda\sigma}}{K_2^2}
\right]\times
\\\nonumber
\left\langle 
\overline{\psi}_{a}\left|\gamma^{\mu} \frac{1}{\slashed{p}-\slashed{K}_1-m-\gamma_{0}V} \gamma^{\lambda}  \frac{1}{\slashed{p}-\slashed{K}_1-\slashed{K}_2-m-\gamma_{0}V} \gamma^{\nu}  \frac{1}{\slashed{p}-\slashed{K}_2-m-\gamma_{0}V} \gamma^{\sigma}\right| \psi_{a}
\right\rangle 
-\delta m^{\mathrm{CL}}.
\end{eqnarray}
\end{widetext}
Decomposition of the second and third electron propagators containing the momentum of the high-energy photon $K_2$ and the accounting of the counter-term $\delta m^{\mathrm{CL}}$, see \cite{Yerokhin:JETP:2005}, yield
\begin{align}
\label{b2reg}
&
\Delta E^{\mathrm{H},\,\mathrm{CL}}_{a}=
\left[-\mathrm{i}e^2\int_{C_1}\frac{d^DK_1}{(2\pi)^D}\frac{g_{\mu\nu}}{K_{1}^2}n_{\beta}(E_{k_1})\right] 
\\\nonumber
&
\times
\left\langle 
\overline{\psi}_{a}\left|
\gamma^{\mu} 
\frac{1}{\slashed{p}-\slashed{K}_1-m-\gamma_{0}V} 
\Gamma^{\mathrm{R}}_{\nu}(p-K_{1},p)
\right| \psi_{a}
\right\rangle 
.
\end{align}

According to the definition of form-factor Eq.~(\ref{b03}), it is easy to find that the square of the transferred momentum $q^2$ in the regularised vertex function %(see Eq.~(\ref{gamR}))
is $q^2=(p-K_{1}-p)^2=K_{1}^2$. As in the case of one-loop self-energy the arguments in the electronic form-factors $F_{1,\,2}$ included in the regularized vertex function can still be represented as an expansions in powers of $q^2$ given by Eqs.~(\ref{b07}) and~(\ref{b07-q}). 
Taking this into account, passing to the Coulomb gauge and coordinate representation, within the nonrelativistic limit in Eq.~(\ref{b2reg}), we arrive at
\begin{widetext}
\begin{eqnarray}
\label{xb2}
\Delta E^{\mathrm{H},\,\mathrm{CL}}_{a}=
e^2\sum\limits_{\pm}\int\limits \frac{d^dk_1}{(2\pi)^d 2k_1}\frac{d-1}{d}n_{\beta}(k_{1})
\left\langle 
\phi_{a}\left|
p_{i} 
\frac{1}{E_{a}-H_{S}\pm k_1} 
\left[
p_{i}\left(-\frac{\alpha}{6\pi\varepsilon}\Delta\right) + \frac{\alpha}{8\pi}\gamma_{0}\sigma_{i j}\nabla^{j}
\right]
\right| \phi_{a}
\right\rangle 
.
\end{eqnarray}
\end{widetext}
Here we used that the operator $q^2$ in the form-factors $F_{1,2}$ acts on a time-independent wave function in the coordinate representation and $q_{\mu}\rightarrow -\mathrm{i}\partial_{\mu}$. 

Then, for $\sigma_{i j}=\varepsilon_{ijk}\Sigma^{k}$ \cite{Greiner}, where $\varepsilon_{ijk}$ is the Levi-Civita antisymmetric tensor and $\Sigma^{k}$ is a component of the Dirac matrix $\bm{\Sigma}$,
%\begin{eqnarray}
%\bm{\Sigma}=
%\begin{pmatrix}
%\bm{\sigma} & 0\\
%0 & \bm{\sigma}
%\end{pmatrix},
%\end{eqnarray}
by introducing the spin operator $\bm{s}=\frac{\bm{\sigma}}{2}$, one can obtain
\begin{eqnarray}
\label{b3}
\Delta E^{\mathrm{H},\,\mathrm{CL}}_{a}=
\frac{2\alpha}{3\pi}\sum\limits_{\pm}\int\limits_{0}^{\infty}dk_{1}k_1 
n_{\beta}(k_{1})\times\qquad
\\\nonumber
\left\langle 
\phi_{a}\left|
p_{i} 
\frac{1}{E_{a}-H_{S}\pm k_1} 
%\times
%\right.
%\right.
%\\\nonumber
%\left.
%\left.
\left[
p_{i}\left(-\frac{\alpha\Delta}{6\pi\varepsilon}\right) - \frac{\alpha}{4\pi}(\bm{s}\times \bm{p})_{i}
\right]
\right| \phi_{a} \right\rangle 
.
\end{eqnarray}
%where the spin operator $\bm{s}=\frac{\bm{\sigma}}{2}$ is introduced.
%Performing the angular integration in the remaining integral, 
Going to the basis set representation and length form in the matrix elements, the finite high-energy part of CL contribution can be reduced to
\begin{eqnarray}
\label{b4}
\Delta E^{\mathrm{H},\,\mathrm{CL}}_{a}=
\frac{2\alpha}{3\pi}
\sum\limits_{\pm}\int\limits_{0}^{\infty} k_{1}^3 n_{\beta}(k_{1})
\left(
\frac{\alpha}{4\pi}
\right)
\\\nonumber
\times
\sum\limits_{n_1}
\frac{\langle 
\phi_{a}
|
r_{i}  | \phi_{n_1} \rangle
\langle 
\phi_{n_1}
|
(\bm{r}\times \bm{s})_{i}
|
\phi_{a}
\rangle }{E_{a}-E_{n_1}\pm k_1} 
.
\end{eqnarray}
The divergent term in Eq. (\ref{b3}) transforms to
\begin{eqnarray}
    \label{divCL}
\frac{2\alpha}{3\pi}
\sum\limits_{\pm}\int\limits_{0}^{\infty} k_{1}^3 n_{\beta}(k_{1})
\left(-\frac{\alpha}{6\pi\varepsilon} \right)
\\
\nonumber
\times
\sum\limits_{n_1}
\frac{\langle 
\phi_{a}
|
r_{i}  | \phi_{n_1} \rangle
\langle 
\phi_{n_1}
|
r_{i}
|
\phi_{n_2}
\rangle
\langle 
\phi_{n_2}
|
\Delta
|
\phi_{n_a}
\rangle
}
{E_{a}-E_{n_1}\pm k_1}.
\end{eqnarray}
%This term is cancelled by the similar one in the low-energy part of CL contribution, see Eq. (\ref{divCL0}). 
This term compensates exactly Eq. (\ref{divCL0}).

\subsubsection{Vacuum loop inside thermal loop}

In this section, we consider the two-loop diagram illustrated in Fig. \ref{fig:5} (b), which represents the case of the ordinary self-energy loop inside the thermal one. The high-energy contribution to the energy shift can be obtained as in \cite{Sapirstein:2004}. For this purpose, we start with the original expression for the energy shift, and use the dimensional regularization. 

In the Feynman gauge we have
\begin{widetext}
\begin{eqnarray}
\label{b01}
\Delta E^{\mathrm{H},\,\mathrm{ViT}}_{a}=-e^4
\left[\int_{C_1} \frac{d^Dk_1}{(2\pi)^D}\frac{g_{\mu\nu}}{K_{1}^2}n_{\beta}(E_{k_1})\right] 
\left[
 \int \frac{d^DK_2}{(2\pi)^D }\frac{g_{\lambda\sigma}}{K_2^2}
\right]
\\\nonumber
\times
\left\langle 
\overline{\psi}_{a}\left|\gamma^{\mu} \frac{1}{\slashed{p}-\slashed{K}_1-m-\gamma_{0}V} \gamma^{\lambda}  \frac{1}{\slashed{p}-\slashed{K}_1-\slashed{K}_2-m-\gamma_{0}V} \gamma^{\nu}  \frac{1}{\slashed{p}-\slashed{K}_1-m-\gamma_{0}V} \gamma^{\sigma}\right| \psi_{a}
\right\rangle 
-\delta m^{\mathrm{ViT}},
\end{eqnarray}
\end{widetext}
where $\delta m^{\mathrm{ViT}}$ gives the mass counter-term. 

The decomposition of the middle electron propagator in Eq.~(\ref{b01}), containing the momentum of the high-energy photon $K_2$, by powers of interaction with the external Coulomb field, leads to
\begin{widetext}
\begin{eqnarray}
\label{b02}
\Delta E^{\mathrm{H},\,\mathrm{ViT}}_{a}=
\left[
-\mathrm{i}e^2\int_{C_1}\frac{d^DK_1}{(2\pi)^D}\frac{g_{\mu\nu}}{K_1^2}n_{\beta}(E_{k_1})
\right] 
% \\\nonumber
% \times
\left[
\left\langle 
\overline{\psi}_{a}\left|\gamma^{\mu} \frac{1}{\slashed{p}-\slashed{K}_1-m-\gamma_{0}V} 
\left[
\Sigma(p')
\right]
\frac{1}{\slashed{p}-\slashed{K}_1-m-\gamma_{0}V} \gamma^{\nu}\right| \psi_{a}
\right\rangle 
\right]
\\\nonumber
+
\left[-
\mathrm{i}e^2\int_{C_1} \frac{d^DK_1}{(2\pi)^D}\frac{g_{\mu\nu}}{K_1^2}n_{\beta}(E_{k_1})
\right] 
% \\\nonumber
% \times
\left\langle 
\overline{\psi}_{a}\left|\gamma^{\mu} 
\frac{1}{\slashed{p}-\slashed{K}_1-m-\gamma_{0}V} 
\left[
\Gamma_{0}(p',p')V
\right]
\frac{1}{\slashed{p}-\slashed{K}_1-m-\gamma_{0}V} \gamma^{\nu}\right| \psi_{a}
\right\rangle 
-\delta m^{\mathrm{ViT}}
,
\end{eqnarray}
\end{widetext}
where $p'=p-K_1$ and $\Gamma_{\mu}(p',p)$ is given by Eq.~(\ref{b03}). 
Then, applying the mass renormalization procedure,%As in the case of one-loop self-energy correction, the mass renormalization leads to the vanishing of the first term in Eq.~(\ref{b02}), and $\Gamma_{\mu}$ should be replaced by the renormalized vertex function $\Gamma^{\mathrm{R}}_{\mu}$:
\begin{widetext}
\begin{eqnarray}
\label{b08}
\Delta E^{\mathrm{H},\,\mathrm{ViT}}_{a}=-\mathrm{i}
e^2\int_{C_1} \frac{d^DK_1}{(2\pi)^D}\frac{g_{\mu\nu}}{K_1^2}n_{\beta}(E_{k_1})
%\\\nonumber
%\times
\left\langle 
\overline{\psi}_{a}
\left|\gamma^{\mu} \frac{1}{\slashed{p}-\slashed{K}_1-m-\gamma_{0}V} 
\right.
\right.
% \\\nonumber
\left[
\Gamma^{\mathrm{R}}_{0}(p',p')V
\right]
%\\\nonumber
%\times
\left.
\left.
\frac{1}{\slashed{p}-\slashed{K}_1-m-\gamma_{0}V} \gamma^{\nu}
\right| \psi_{a}
\right\rangle. \qquad
\end{eqnarray}
\end{widetext}

To identify the divergent contribution with similar one in the low energy part we turn to the Coulomb gauge and nonrelativistic limit in the remaining matrix elements of Eq.~(\ref{b08}). As in the case of the one thermal loop, the integration over $k_{10}$ does not involve the poles of the electron propagators. Then,
\begin{align}
\label{asd4}
&
\Delta E^{\mathrm{H},\,\mathrm{ViT}}_{a}=
e^2\sum\limits_{\pm}\int\limits \frac{d^dk_1}{(2\pi)^d 2k_1}\frac{d-1}{d}n_{\beta}(k_{1})
\\\nonumber
&
\times
\left\langle 
\phi_{a}
\left|p_{i} \frac{1}{E_{a}-H_{S}\pm k_{1}} 
\right.
\right.
% \\\nonumber
% &
% \times
\left[
\Gamma^{\mathrm{R}}_{0}V
\right]
%\\\nonumber
%\times
\left.
\left.
\frac{1}{E_{a}-H_{S}\pm k_{1}} p_{i}
\right| \phi_{a}
\right\rangle. 
\end{align}
%In the basis set representation and length form of matrix element Eq.~(\ref{asd4}) takes the form
In the basis set representation and length form:
\begin{align}
\label{asd5}
&
\Delta E^{\mathrm{H},\,\mathrm{ViT}}_{a}=
\frac{2\alpha}{3\pi}\int\limits_{0}^{\infty} dk_{1}k_{1}^3n_{\beta}(k_{1})
\sum\limits_{\pm}
\sum\limits_{n_1n_2}
\\\nonumber
\times
&
\frac{
\langle 
\phi_{a}
|r_{i} |\phi_{n_{1}}\rangle \langle \phi_{n_{1}}|\Gamma^{\mathrm{R}}_{0}V|\phi_{n_{3}}\rangle_{\mathrm{reg}} \langle \phi_{n_{3}}|r_{i}| \phi_{a}
\rangle}{(E_{a}-E_{n_1}\pm k_{1})(E_{a}-E_{n_3}\pm k_{1})},
\end{align}   
where the matrix element of the operator $\Gamma^{\mathrm{R}}_{0}V$ is given by Eq.~(\ref{GVreg}). 
The singular term omitted in Eq.~(\ref{asd5}) is
\begin{align}
&
\frac{2\alpha}{3\pi}\int\limits_{0}^{\infty} dk_{1}k_{1}^3n_{\beta}(k_{1})
\sum\limits_{\pm}
\sum\limits_{n_1n_2}
\left(-\frac{\alpha}{6\pi\varepsilon}\right)
\\\nonumber
&
\times
\frac{
\langle 
\phi_{a}
|r_{i} |\phi_{n_{1}}\rangle \langle \phi_{n_{1}}|\Delta V|\phi_{n_{3}}\rangle \langle \phi_{n_{3}}|r_{i}| \phi_{a}
\rangle}{(E_{a}-E_{n_1}\pm k_{1})(E_{a}-E_{n_3}\pm k_{1})}  
,
\end{align}
and is exactly reduced by the divergent term in the low-energy part of the ViT contribution.

\subsubsection{Vacuum loop over thermal loop}

In a similar way, for the VoT contribution in the Feynman gauge, one can write
\begin{widetext}
\begin{eqnarray}
\label{b091}
\Delta E^{\mathrm{H},\,\mathrm{VoT}}_{a}=-e^4
\left[\int_{C_1} \frac{d^Dk_1}{(2\pi)^D}\frac{g_{\mu\nu}}{K_{1}^2}n_{\beta}(E_{k_1})\right] 
\left[
 \int \frac{d^DK_2}{(2\pi)^D }\frac{g_{\lambda\sigma}}{K_2^2}
\right]
\\\nonumber
\times
\left\langle 
\overline{\psi}_{a}\left|\gamma^{\mu} \frac{1}{\slashed{p}-\slashed{K}_2-m-\gamma_{0}V} \gamma^{\lambda}  \frac{1}{\slashed{p}-\slashed{K}_1-\slashed{K}_2-m-\gamma_{0}V} \gamma^{\sigma}  \frac{1}{\slashed{p}-\slashed{K}_2-m-\gamma_{0}V} \gamma^{\nu}\right| \psi_{a}
\right\rangle 
-\delta m^{\mathrm{VoT}}
\end{eqnarray}
\end{widetext}
with $\delta m^{\mathrm{VoT}}$ representing the mass counter-term. 

The divergences in the high-energy part of the CL, ViT, and LaL diagrams can be separated by decomposing the electron propagators into powers of interaction with the Coulomb field of the nucleus and then forming a vertex function. In the case of the VoT contribution, the separation of the divergent terms in the original representation, Eq. (\ref{b091}), becomes less evident due to the presence of three denominators with $K_2$ momentum. Considering the complications of the overall mathematical calculations, we focus on describing the regularization procedure in general. The first step involves the decomposition of all three propagators into powers of the Coulomb interaction. Then only those contributions consisting of a single factor $\gamma_{0}V$ sandwiched between the free electron propagators are considered. Angular algebra is then performed using gamma matrices in Eq. (\ref{b091}), followed by application of the Feynman parametrization and dimensional integration.

\subsubsection{Loop after loop}

As mentioned above, the contribution of two-loop LaL diagrams (see graphs (c) and (d) in Fig.~\ref{fig:5}) %to the atomic state energy shift
can be partitioned into irreducible and reducible parts:
\begin{eqnarray}
%&
\Delta E^{\mathrm{H},\,\mathrm{LaL}}_{a}=\Delta E^{\mathrm{H},\,\mathrm{LaL}}_{a, \mathrm{irr}}+\Delta E^{\mathrm{H},\,\mathrm{LaL}}_{a, \mathrm{red}}
%\\\nonumber
%&
- \delta m^{\mathrm{LaL}}
,\qquad
\end{eqnarray}
%where
\begin{eqnarray}
%\\
\label{b21}
\Delta E^{\mathrm{LaL}}_{a, \mathrm{irr}}= 
\left[e^2\int_{C_1} \frac{d^DK_1}{(2\pi)^D}\frac{g_{\mu\lambda}}{K_1^2}n_{\beta}(E_{k_1})\right]\times\qquad
\\\nonumber
\left[
e^2 \int \frac{d^DK_2}{(2\pi)^D }\frac{g_{\nu\sigma}}{K_2^2}
\right]
\left\langle 
\overline{\psi}_{a}\left|\gamma^{\mu} \frac{1}{\slashed{p}-\slashed{K}_1-m-\gamma_{0}V} \gamma^{\lambda}\times
\right.\right.
\\\nonumber
\left.\left.
\frac{1}{(\slashed{p}-m-\gamma_{0}V)'} \gamma^{\nu}  \frac{1}{\slashed{p}-\slashed{K}_2-m-\gamma_{0}V} \gamma^{\sigma}\right| \psi_{a}
\right\rangle 
,
\end{eqnarray}
%and
\begin{eqnarray}
%\\
\label{b22}
%&
\Delta E^{\mathrm{LaL}}_{a, \mathrm{red}}
\equiv \Delta E^{\mathrm{H},\,\mathrm{LaL}}_{a, \mathrm{red}\,1} +  \Delta E^{\mathrm{H},\,\mathrm{LaL}}_{a, \mathrm{red}\,2} = \qquad
\\\nonumber
%&
-
\frac{1}{2}
\left[e^2\int_{C_1} \frac{d^DK_1}{(2\pi)^D}\frac{g_{\mu\lambda}}{K_1^2}n_{\beta}(E_{k_1})\right] 
\left[
e^2 \int\frac{d^D K_2}{(2\pi)^D }\frac{g_{\nu\sigma}}{K_2^2}
\right]
\\\nonumber
%&
\times
\left(
\left\langle 
\overline{\psi}_{a}\left|\gamma^{\mu} \frac{1}{\slashed{p}-\slashed{K}_1-m-\gamma_{0}V} \gamma^{\lambda} 
\right| \psi_{a}
\right\rangle 
\right.
\\\nonumber
%&
\times
\frac{\partial}{\partial E_{a}}
\left\langle 
\overline{\psi}_{a}\left|
\gamma^{\nu}
\frac{1}{\slashed{p}-\slashed{K}_2-m-\gamma_{0}V} \gamma^{\sigma}\right| \psi_{a}
\right\rangle 
\\\nonumber
%&
+
\left\langle 
\overline{\psi}_{a}\left|
\gamma^{\nu}
\frac{1}{\slashed{p}-\slashed{K}_2-m-\gamma_{0}V} \gamma^{\sigma}\right| \psi_{a}
\right\rangle 
\\\nonumber
%&
\times
\left.
\frac{\partial}{\partial E_{a}}
\left\langle 
\overline{\psi}_{a}\left|\gamma^{\mu} \frac{1}{\slashed{p}-\slashed{K}_1-m-\gamma_{0}V} \gamma^{\lambda} 
\right| \psi_{a}
\right\rangle 
\right).
\end{eqnarray}

In complete analogy with the evaluation of CL, ViT and VoT, the finite part of the irreducible high energy LaL contribution reduces to %in the nonrelativistic limit is reduced to the expression:
\begin{align}
\label{irr_nr}
&
\Delta E^{\mathrm{H},\,\mathrm{LaL}}_{a, \mathrm{irr}}=
\frac{2\alpha}{3\pi}\int\limits_{0}^{\infty} dk_{1}k_{1}^3n_{\beta}(k_{1})
% \\\nonumber
% \times
\sum\limits_{\pm}
\sum\limits_{\substack{n_1\\n_2\neq a}}
\\\nonumber
&
\times
\frac{
\langle 
\phi_{a}
|r_{i} |n_{1}\rangle
\langle \phi_{n_{1}}|r_{i}| \phi_{n_2}
\rangle
\langle \phi_{n_{2}}|\Gamma^{\mathrm{R}}_{0}V| \phi_{a}
\rangle_{\mathrm{reg}}}{(E_{a}-E_{n_1}\pm k_{1})(E_{a}-E_{n_2})} 
.
\end{align}  

As can be seen from Eq.~(\ref{b22}), the reducible part of loop-after-loop diagrams is represented by the product of two diagonal matrix elements of finite temperature SE and energy derivative of ordinary SE operator and vice versa. Therefore, to calculate this contribution, one can directly use the regularized one-loop results for low and high energy parts $\Delta E^{\mathrm{L+H},\,\mathrm{LaL}}_{a}$. Then, according to sections~\ref{section:1} and~\ref{section:2}, the reducible part of loop-after-loop diagrams is given by 
\begin{widetext}
\begin{eqnarray}  
\label{red_f1}
 \Delta E^{\mathrm{L+H},\,\mathrm{LaL}}_{a, \mathrm{red}\,1} = 
 \frac{1}{2}
 \left[
 \frac{2\alpha}{3\pi}\sum\limits_{\pm}\sum\limits_{n_1}\int\limits_{0}^{\infty}  dk_1\, k_1^3\, n_{\beta}(k_1)
% \right.
% \qquad
%\\\nonumber
%\times
%\left.
\frac{\langle \phi_{a} |r_{i}| \phi_{n_1}\rangle \langle \phi_{n_1} |r_{i}|\phi_{a}\rangle}
{
E_a-E_{n_1}\pm k_1
}
\right]
% \\\nonumber
% \times
\frac{\partial}{\partial E_{a}}
 \left[
 \frac{4\alpha (\alpha Z)^4}{3\pi n_{a}^3}\log\beta_{a}
 \right]
 ,
\end{eqnarray}
\begin{eqnarray} 
\label{red_f2}
 \Delta E^{\mathrm{L+H}\,,\mathrm{LaL}}_{a, \mathrm{red}\,2} = 
 -
 \frac{1}{2}
 \left[
\frac{4\alpha (\alpha Z)^4}{3\pi n_{a}^3}
\left[
\left(
\frac{5}{6}
-
2\log (\alpha Z)
\right)
\delta_{l_{a}0}
- 
\log \beta_{a}
\right]
% \right.
% \\\nonumber
% \left.
-\frac{\alpha(Z\alpha)^4 }{2\pi n_{a}^3} \frac{\left(j_{a} (j_{a}+1)-l_{a}(l_{a}+1)-\frac{3}{4}\right)}{l_{a}(l_{a}+1)(2l_{a}+1)}
\right]
\\\nonumber
 \times
 \left[
\frac{2\alpha}{3\pi}\sum\limits_{\pm}\sum\limits_{n_3}\int\limits_{0}^{\infty}  dk_1\, k_1^3\, n_{\beta}(k_1)
% \right.
%  \qquad
% \\\nonumber
% \times
% \left.
\frac{\langle \phi_{a} |r_{i}| \phi_{n_3}\rangle \langle \phi_{n_3} |r_{i}|\phi_{a}\rangle}
{
(E_a-E_{n_3}\pm k_1)^2
}
\right]
.
\end{eqnarray}
\end{widetext}

The final formulas obtained for the combined two-loop diagrams are given by the expressions Eqs. (\ref{finiteCL}), (\ref{b009}), (\ref{lal3}), (\ref{b4}), (\ref{asd4}) and (\ref{irr_nr})-(\ref{red_f2}), which can be further evaluated numerically. %The main difficulty is the triple summation over the entire spectrum of intermediate states. To solve this problem, as in the case of a one-loop thermal shift, we will use the B-spline approach for the solution of Schr\"odinger equation for hydrogen atom \cite{DKB}, which leads to summation over the discrete spectrum of pseudostates.
To perform a triple summation over the entire spectrum of intermediate states, we further use the B-splines method \cite{DKB}, which restores solutions of the Schr\"{o}dinger equation by replacing the entire spectrum of states onto a finite number of pseudo-states.

\section{Results and conclusions}
\label{final}

In this work, we focus exclusively on calculating the combined two-loop self-energy corrections for the low-lying $s$-states in the hydrogen atom. The combination of two-loop diagrams is represented by one thermal and one loop pertaining to the zero temperature case, see Fig.~\ref{fig:5}. The final results are given by the expressions in Eqs. (\ref{finiteCL}), (\ref{b009}), (\ref{lal3}), (\ref{b4}), (\ref{asd4}) and (\ref{irr_nr})-(\ref{red_f2}). As it was found, the high-energy contribution vanishes upon integrating over the angles in the relevant matrix elements, just as it does in the case of one-loop self-energy correction. Hence, for $s$-states, all computations reduce to a numerical calculation of the low-energy part only.

The primary challenge in performing numerical calculations of higher-order QED corrections lies in computing sums over intermediate states. We utilize B-splines method to approximate the solutions of the Schr\"{o}dinger equation for the hydrogen atom and generate an appropriate basis set. This enables us to replace the sums over the intermediate states, which include the continuum spectrum, with sums over the discrete set of pseudo-states. We are interested in calculating two-loop radiative corrections for the ground $1s$ and metastable $2s$ states in the hydrogen atom due to their particular importance in modern spectroscopic experiments, see, e.g., \cite{Rothery,Fischer,Parthey:2s1s}. Thus, we generated basis sets on $ns$, $np$, and $nd$ intermediate states, each of which is $n\leq 40$ long, letting us verify convergence and guarantee the digits presented in all obtained results.

Table \ref{tab:1} provides the numerical values of the combined two-loop self-energy corrections for two different temperatures, namely $T=300$ and $T=1000$ K. It is worth noting that the magnitude of the calculated corrections aligns with the rough estimates presented in the recent study \cite{chinese_twoloop}. These corrections are also comparable to the thermal corrections to the Breit interactions discussed earlier in \cite{SZA_2021}.
\begin{table}[ht]
        \caption{Combined two-loop self-energy corrections for the $1s$ and $2s$ states of hydrogen atom at different temperatures (in Kelvin). All values are in Hz.}
            \label{tab:1}
    \centering
    \begin{tabular}{l c c}

      \hline
      \hline
             & $T=300$ K & \\
      \hline
                                     & $1s$ & $2s$ \\
      \hline
      $\Delta E_{a}^{\mathrm{CL}}$   & $-1.59\times 10^{-7}$ & $-2.13\times 10^{-6}$ \\ 
      $\Delta E_{a}^{\mathrm{ViT}}$  & $1.57\times 10^{-7}$ & $3.00\times 10^{-5}$\\
      $\Delta E_{a}^{\mathrm{VoT}} $ & $-1.28\times 10^{-4}$ & $ -9.69\times 10^{-5} $\\
      $\Delta E_{a, \mathrm{irr}}^{\mathrm{LaL}}$ & $-1.37\times 10^{-7}$& $ -4.11\times 10^{-6} $\\
      $\Delta E_{a, \mathrm{red}\,1}^{\mathrm{LaL}}$ & $3.49\times 10^{-8}$& $3.50\times 10^{-7}$\\
      $\Delta E_{a, \mathrm{red}\,2}^{\mathrm{LaL}}$ &$-5.87\times 10^{-8}$ &$-3.53\times 10^{-3}$ \\
      $\Delta E_{a}^{\mathbf{total}}$ & $\bm{-1.28\times 10^{-4}}$ & $ \bm{-3.53\times 10^{-3}} $\\
      \hline
      \hline
             & $T=1000$ K & \\
      \hline
                                     & $1s$ & $2s$ \\
      \hline
      $\Delta E_{a}^{\mathrm{CL}}$   & $-1.97\times 10^{-5}$ & $-2.64\times 10^{-4}$\\
      $\Delta E_{a}^{\mathrm{ViT}}$  & $1.96\times 10^{-5}$ & $7.46\times 10^{-4}$\\
      $\Delta E_{a}^{\mathrm{VoT}} $ & $-1.17\times 10^{-2}$ & $-9.13\times 10^{-3}$\\
      $\Delta E_{a,\mathrm{irr}}^{\mathrm{LaL}}$ & $-1.69\times 10^{-5}$ & $ -5.26\times 10^{-4}$\\
      $\Delta E_{a,\mathrm{red}\,1}^{\mathrm{LaL}}$ & $4.31\times 10^{-6}$ & $4.46\times 10^{-5}$\\
      $\Delta E_{a, \mathrm{red}\,2}^{\mathrm{LaL}}$ & $-7.27\times 10^{-6}$ & $-3.93\times 10^{-2}$ \\
      $\Delta E_{a}^{\mathbf{total}}$ & $\bm{-1.17\times 10^{-2}}$ & $\bm{-4.84\times 10^{-2}}$\\
      \hline
    \end{tabular}
\end{table}

As can be seen from Table \ref{tab:1}, the largest contribution comes from the reducible part of the loop-after-loop diagram, it is the correction defined by Eq. (\ref{red_f2}). Fortunately, the latter is the easiest expression to calculate among all those we have considered in this work. The value of $\Delta E_{a, \mathrm{red}\,2}^{\mathrm{LaL}}$, which reaches $10^{-3}$ level in order of magnitude at $T=300$ K, demonstrates the role of the effect considered here. It can be straightforwardly compared with the well-known Stark shift caused by the blackbody radiation \cite{Farley}. For the hydrogen atom in the BBR environment at $300$ K, the AC-Stark shift is about $0.04$ Hz and $1$ Hz for $1s$ and $2s$ states, respectively, see \cite{Farley,Jentschura:BBR:2008,SLP-QED}. In the application to hydrogen spectroscopy, the combined two-loop corrections is of the same order as other effects contributing to the uncertainty budget in measuring the $1s-2s$ transition frequency, which stands at 10 Hz (see Table I in \cite{Parthey:2s1s}). As a consequence, these effect can be relevant even for the current experimental setup. However, one can expect that with a subsequent increase in measurement accuracy, the considered correction may gain importance. Furthermore, it is essential to conduct a separate analysis for highly excited states of the hydrogen atom since thermal contributions increase with the principal quantum number.

%It should be noted that the considered combined two-loop correction may be also important for the measurement of the hyperfine structure (HFS) of the ground state of the hydrogen atom which was carried out with an experimental error $10^{-3}$ Hz \cite{Hellwig}. This is just several times larger than the effect found in our paper (it is $10^{-4}$ at room temperature). In addition, it can be expected that the total contribution of the combined two-loop diagrams can exceed the corrections to the hyperfine structure in the hydrogen atom considered in the work \cite{Volotka2005}. Although the relevant calculations for HFS are beyond the scope of the present work, it can be expected that the effect may be significant. 

Table \ref{tab:1} also shows the values of two-loop self-energy corrections calculated at a temperature of $1000$ K. A direct comparison of numerical results for two different temperatures leads to the conclusion that the parametric estimates on temperature given, for example, in \cite{chinese_twoloop}, are hardly applicable to the two-loop contributions. In contrast to the one-loop contribution, this follows from the integration over the frequencies involving the Planck distribution function. For example, the correction $\Delta E_{a, \mathrm{red}\,2}^{\mathrm{LaL}}$ does not scale as $T^4$ or similar. Thus, numerical calculations should be performed to obtain an accurate value for each specific temperature.

Along with the fundamental generalization of the TQED theory for bound states to the two-loop level presented in this paper, one can draw attention to the following. The corrections given by expressions Eqs. (\ref{finiteCL}), (\ref{b009}), (\ref{lal3}), (\ref{b4}), (\ref{asd4}) and (\ref{irr_nr})-(\ref{red_f2}) can be considered in the context of precision experiments carried out with systems relating to atomic clocks \cite{al+obs,baynham}. 

In the examination of two-loop corrections in systems such as the hydrogen atom, our analysis reveals that their impact on the transition frequency determination in spectroscopic measurements  can only make a small contributions the uncertainty budget on the level of recoil effect \cite{Parthey:2s1s}.  Nonetheless, it is worth noting that when dealing with highly excited Rydberg states and extreme astrophysical conditions characterized by elevated temperatures, it is reasonable to conjecture that the aforementioned correction may assume a more substantial role. Thus, the investigation of this particular matter warrants a separate and focused study.

It is important to note that the estimates of the effect for neutral atoms closely align with those for hydrogen in terms of the same order of magnitude. As a particular example, one can consider alkali group atoms, where the thermal Stark shifts of atomic levels (thermal one-loop QED correction) are of the same order as in neutral hydrogen \cite{Farley}. At the same time, the Lamb shift of the ground and first excited states (due to zero-temperature QED corrections from one loop and the vacuum polarization contribution) can be one or two orders higher than in hydrogen \cite{lamb1, lamb2, lamb3}. Since the combined correction is proportional to the product of zero and finite-temperature one-loop contributions, it is reasonable to expect that in many-electron atoms, the shift, even for the ground state, will be at least of the same order or even greater than $10^{-4}$ Hz. Moreover, the effects of the order $10^{-3}$-$10^{-4}$ Hz correspond to dynamic corrections to the thermal Stark shift in Ca$^{+}$ and Sr$^{+}$ atomic clocks \cite{Sarfonova_2012}. Recently, dynamic corrections to the BBR shift at the level of $10^{-5}$ Hz have also been discussed in connection with molecular Sr$_{2}$ clocks (see Table II in \cite{iritani2023precise}). In addition, when considering the Yb atom, which also serves as the foundation for some of the most accurate atomic clocks, the thermal Stark shift has been meticulously measured in \cite{Beloy} and found to be $-1.25487(2)$ Hz. The associated error, $10^{-5}$, is at the same level as the effect observed for the ground state. This contribution arises due to the corresponding symmetrization of the wave function of a system of two atoms and appears only if atoms of the same type are considered in different states. This problem was covered in detail in a series of articles

%This leads to a rather interesting effect, when the mixed correction arising in the second order of the perturbation theory and contributing to the total energy difference $2s-1s$ can almost reach the order of magnitude as the pure thermal contribution in the first order (i.e., it becomes comparable to the ordinary BBR-induced Stark). Therefore, we can conclude that the effect being considered may also have a significant impact on atomic clocks that utilize the ground and first excited states as clock transition. Furthermore, the current accuracy achieved in measuring the $1s-2s$ transition frequency, which is 10 Hz \cite{Parthey:2s1s}, enables the observation of thermal effects of higher orders. Based on the findings of this study, such an experiment could be conducted in the presence of an additional source of black-body radiation, heated to $1000$ degrees Kelvin or higher, which is readily attainable even in laboratory settings.

\section{Acknowledgements}
This work was supported by President grant MK-4796.2022.1.2 and foundation for the advancement of mathematics and theoretical physics "BASIS" (grant No. 23-1-3-31-1). Evaluation of high energy contributions (section~\ref{HE}) was supported by the Russian Science Foundation under grant No. 22-12-00043.

\appendix
\renewcommand{\theequation}{A\arabic{equation}}
\setcounter{equation}{0}

\section{Dimensional regularization of loop integrals}
\label{appendix:A}

This Appendix contains the results of evaluating $d$-dimensional integrals over photon momentum in one- and two-loop contributions. The first step involves performing spatial integration. In $d=3-2\varepsilon$ dimensions this can be accomplished using Eq. (\ref{dint}) provided in the main text. %Subsequently, the remaining integral along the real half-axis can be conveniently computed by applying the Cauchy theorem, and the outcome can be expanded as a Taylor series in powers of $\varepsilon$.
The remaining integral along the real half-axis can be calculated by applying Cauchy theorem, and then the result can be expanded into a Taylor series of $\varepsilon$ powers.

%Here, we present the final result that arises from calculating the one-loop self-energy correction, which involves the following integral:
The following integral is involved in calculating the one-loop self-energy correction:
\begin{eqnarray}
\label{a1}
I_{1}=
e^2
\int \frac{d^dk}{(2\pi)^d 2k}
\left(
\delta_{ij}-\frac{k_ik_j}{\bm{k}^2}
\right)
p_{i}
\frac{1}{E_{a}-H_{S}-k_2}
p_{j}
\qquad
&
\\\nonumber
=e^2
\frac{2\pi^{d/2}}{\Gamma(d/2)}\frac{d-1}{d}\frac{1}{2(2\pi)^d}
\int\limits_{0}^{\infty}k^{d-2}  dk
\,
p_{i}
\frac{1}{E_{a}-H_{S}-k}
p_{i}
&
\\\nonumber
\approx
\frac{2\alpha}{3 \pi } 
p_{i}
(H_{S}-E_{a})
\left(
\frac{1}{2\varepsilon}-\log [2(H_{S}-E_{a})] +\frac{5}{6} -\frac{\gamma_{\mathrm{E}}}{2} 
\right.
&
\\\nonumber
\left.
+\frac{1}{2}\log 4\pi
\right)
p_{i}
.
\end{eqnarray}
As can be seen from this expression, all divergences arising in the loop diagrams are explicitly concatenated with the division by the parameter $\varepsilon\rightarrow 0$. In the spectral representation of electron propagator this integral can be written as
\begin{align}
&
I_{1}=   
\frac{2\alpha}{3 \pi } \sum\limits_{n}
p_{i}|n\rangle
\langle  n | p_{i}
(E_{n}-E_{a})
\\\nonumber
&
\times
\left(
\frac{1}{2\varepsilon}-\log [2|E_{n}-E_{a}|] +\frac{5}{6} -\frac{\gamma_{\mathrm{E}}}{2} 
% \right.
% &
% \\\nonumber
% \left.
+\frac{1}{2}\log 4\pi
\right)
.
\end{align}

A similar integral can be taken in the length gauge, which gives an additional $k^2$ factor in the numerator of the integrand in the Eq.~(\ref{a1}):
\label{appA}
\begin{align}
\label{a1length}
&
I_{1}=
e^2
\int \frac{d^dk\,k^2}{(2\pi)^d 2k}
\left(
\delta_{ij}-\frac{k_ik_j}{\bm{k}^2}
\right)
r_{i}
\frac{1}{E_{a}-H_{S}-k_2}
r_{j}
\\\nonumber
&
=e^2
\frac{2\pi^{d/2}}{\Gamma(d/2)}\frac{d-1}{d}\frac{1}{2(2\pi)^d}
\int\limits_{0}^{\infty}k^{d}   dk
\,
r_{i}
\frac{1}{E_{a}-H_{S}-k_2}
r_{i}
\\\nonumber
&
\approx
\frac{2\alpha}{3 \pi } 
r_{i}
(H_{S}-E_{a})^3
\left(
\frac{1}{2\varepsilon}-\log [2(H_{S}-E_{a})] +\frac{5}{6} 
\right.
\\\nonumber
&
\left.
-\frac{\gamma_{\mathrm{E}}}{2} 
+\frac{1}{2}\log 4\pi
\right)
r_{i}.
\end{align}
In the spectral representation of electron propagator one can obtain
\begin{align}
&
 I_{1}=   
 \frac{2\alpha}{3 \pi } \sum\limits_{n}
r_{i}|n\rangle
\langle  n | r_{i}
(E_{n}-E_{a})^3
\\\nonumber
&
\times
\left(
\frac{1}{2\varepsilon}-\log [2|E_{n}-E_{a}|] +\frac{5}{6} -\frac{\gamma_{\mathrm{E}}}{2} 
% \right.
% &
% \\\nonumber
% \left.
+\frac{1}{2}\log 4\pi
\right)
.
\end{align}

The integral arising in the reducible part of loop-after-loop diagrams can  be easily evaluated as
\begin{eqnarray}
\label{a2}
%&
I_{2}=-\frac{\partial}{\partial E_{a}}I_{1}=
e^2\int \frac{d^dk \,k^2}{(2\pi)^d 2k}
\left(
\delta_{ij}-\frac{k_ik_j}{\bm{k}^2}
\right)\qquad
\\\nonumber
%&
\times
r_{i}
\frac{1}{(E_{a}-H_{S}-k)^2}
r_{j}
\approx
\frac{2\alpha}{3 \pi }
r_{i}(3(H_{S}-E_{a})^2)
\\\nonumber
%&
\left(\frac{1}{2\varepsilon}-\log[2 (H_{S}-E_{a})]+\frac{1}{2}-\frac{ \gamma_{\mathrm{E}}}{2} 
+\frac{1}{2}\log 4\pi 
\right)
r_{i},
\end{eqnarray}
and similarly for spectral representation
\begin{eqnarray}
% &
 I_{2}=   
 \frac{2\alpha}{3 \pi } \sum\limits_{n}
r_{i}|n\rangle
\langle  n | r_{i}
(3(E_{n}-E_{a})^2)\qquad
\\\nonumber
\times
%&
\left(
\frac{1}{2\varepsilon}-\log [2|E_{n}-E_{a}|] +\frac{1}{2} -\frac{\gamma_{\mathrm{E}}}{2} 
% \right.
% &
% \\\nonumber
% \left.
+\frac{1}{2}\log 4\pi
\right)
.
\end{eqnarray}

The corresponding integral arising for crossed loops diagrams in the spectral representation of electron propagators can be calculated found in the form:
\begin{widetext}
\begin{eqnarray}
\label{a3}
%&
I_{3}=
e^2\int \frac{d^dk_2\,k_2^2}{(2\pi)^d 2k_2}
\left(
\delta_{ij}-\frac{k_ik_j}{\bm{k}^2}
\right)
r_{i}
\frac{1}{(E_{a}-H_{S}\pm k_1 -k_2)}r_{k}
\frac{1}{(E_{a}-H_{S}-k_2)}
r_{j}
= \qquad
 \\\nonumber
%&
%\times
\frac{2\alpha}{3\pi}\sum\limits_{\pm}\sum\limits_{n_2n_3}
r_{i}|\phi_{n_{2}} \rangle\langle \phi_{n_{2}}| r_{k}|\phi_{n_{3}}\rangle  \langle \phi_{n_{3}}|r_{i}
\left[
\left(
\left(E_{a}-E_{n_2}\pm k_1\right)^2+ 
(E_{a}-E_{n_2}\pm k_1)(E_{a}-E_{n_3})+
\left(E_{a}-E_{n_3}\right)^2 \right) \times
% \\\nonumber
% &
% \times
\right.
\\\nonumber
\left.
%&
\left(
\frac{1}{2\varepsilon}
+
\frac{5}{6}
-
\frac{\gamma_{\mathrm{E}}}{2}
+\frac{1}{2}\log 4\pi
\right)
-
\frac{
(E_{a}-E_{n_2}\pm k_1)^3
\log[2|E_{a}-E_{n_2}\pm k_1|]}
{E_{n_3}-E_{n_2}\pm k_1} 
% \\\nonumber
% &
%\left.
+
\frac{
(E_{a}-E_{n_3})^3 \log[2|E_{a}-E_{n_3}|]}
{E_{n_3}-E_{n_2}\pm k_1}
\right]
.
\end{eqnarray}
%\end{widetext}

%\begin{widetext}
\begin{eqnarray}
\label{a4}
%&
I_{4}=
e^2\int \frac{d^dk_2\,k_2^2}{(2\pi)^d 2k_2}
\left(
\delta_{ij}-\frac{k_ik_j}{\bm{k}^2}
\right)
r_{i}
\frac{1}{(E_{a}-H_{S}-k_2)}
r_{k}
\frac{1}{(E_{a}-H_{S}\pm k_1 -k_2)}
r_{l}
\frac{1}{(E_{a}-H_{S}-k_2)}
r_{j}
 \\\nonumber
 =
\frac{2\alpha}{3\pi}\sum\limits_{\pm}\sum\limits_{n_1n_2n_3}
r_{i}|\phi_{n_{1}} \rangle
\langle \phi_{n_{2}}| r_{k}|\phi_{n_{2}}\rangle 
\langle \phi_{n_{2}}| r_{l}|\phi_{n_{3}}\rangle  \langle \phi_{n_{3}}|r_{i}
 \\\nonumber
\times
\left[
(3 E_{a}-E_{n_1}-E_{n_2}-E_{n_3}\pm k_{1})
\left(
\frac{(E_{a}-E_{n_1})^3 \log [2
   |E_{a}-E_{n_1}|]-(E_{a}-E_{n_2}\pm k_{1})^3 \log [2
   |E_{a}-E_{n_2}\pm k_{1}|]}{(E_{n_1}-E_{n_2}\pm k_{1}) (-E_{n_2}+E_{n_3}\pm k_{1}) (3
   E_{a}-E_{n_1}-E_{n_2}-E_{n_3}\pm k_{1})}
   \right.
   \right.
   \\\nonumber
   \left.
   \left.
   -\frac{(E_{a}-E_{n_1})^3 
   \log [2    |E_{a}-E_{n_1}|]-(E_{a}-E_{n_3})^3 \log [2 |E_{a}-E_{n_3}|]}{(E_{n_1}-E_{n_3})
   (-E_{n_2}+E_{n_3}\pm k_{1}) (3 E_{a}-E_{n_1}-E_{n_2}-E_{n_3}\pm k_{1})}+\frac{1}{2
   \varepsilon}-\frac{\gamma_{\mathrm{E}} }{2}+\frac{5}{6}+\frac{1}{2} \log (4 \pi )
   \right)
\right]
.
\end{eqnarray}
\end{widetext}

\renewcommand{\theequation}{B\arabic{equation}}
\setcounter{equation}{0}

\section{Relations between matrix elements in velocity and length form}
\label{appendix:B}

In order to represent the arising expressions in a form convenient for numerical calculations, the transformation of matrix elements in Eq. (\ref{xbeta7}) to the length gauge should be performed. With the use of commutation relations
\begin{eqnarray}
\label{commutator}
  p_{i}=\mathrm{i}[H_{S},r_{i}]  =\mathrm{i}[H_{S}-E + k,r_{i}],  
\end{eqnarray} 
it is easy to show that the following equality is valid 
\begin{align}
\label{aa1}
&
    \left\langle \phi_{n'}\left|
    p_{i}
    \frac{1}{E_{n'}-H_{S}-k}
    p_{j}
   \right| \phi_{n} \right\rangle
   =     
    \\\nonumber
&    
    -k(E_{n'} - E_n -k)
    \left\langle \phi_{n'}\left|
    r_{i}
    \frac{1}{E_{n'}-H_{S}-k}
    r_{j}
   \right| \phi_{n} \right\rangle       \qquad  
         \\\nonumber
&         
         +
         (k-\frac{1}{2}(E_{n'}-E_{n}))
          \left\langle \phi_{n'}\left|
   r_{i}r_{j}
   \right| \phi_{n} \right\rangle
   +\frac{3}{2}\delta_{n'n}
   .\qquad
\end{align}
In the case of $n=n'=a$, Eq.~(\ref{aa1}) turns to
\begin{eqnarray}
\label{aa2}
\sum\limits_{\pm}
    \left\langle \phi_{a}\left|
    p_{i}
    \frac{1}{E_{a}-H_{S} \pm k}
    p_{i}
   \right| \phi_{a} \right\rangle
   \\\nonumber
   =
   k^2
   \sum\limits_{\pm}
   \left\langle \phi_{a}\left|
    r_{i}
    \frac{1}{E_{a}-H_{S} \pm k}
    r_{i}
   \right| \phi_{a} \right\rangle + 3 
.   
\end{eqnarray}

As shown in the main text, the gauge transformation, performed using Eqs. (\ref{aa1}) and (\ref{aa2}), leads to the natural separation of the mass counter-term and correct temperature behavior when making parametric estimations.

\renewcommand{\theequation}{C\arabic{equation}}
\setcounter{equation}{0}

\section{Evaluation of matrix elements and commutation relations}
\label{appendix:C}
Cyclic component of vector product of position and spin operators in the matrix elements of Eq.~(\ref{b4}) can be written in terms of tensor product as follows
\begin{eqnarray}
  (\bm{r}\times \bm{s})_{1q}= - \mathrm{i}\sqrt{2}\lbrace s_{1}\otimes r_{1} \rbrace_{1q}  
\end{eqnarray}
Then with the use of standard technique for the evaluation of matrix elements we find \cite{VMK}
\begin{eqnarray}
\langle n'l's'j'm' | (\bm{r}\times \bm{s})_{1q} | nlsjm\rangle = \qquad
\\\nonumber
%\times
-
\mathrm{i}\sqrt{6} 
(-1)^{j'-m'}\sqrt{(2j'+1)(2j+1)}\times
\\\nonumber
\begin{pmatrix}
    j' & 1 & j \\
   -m' & q & m
\end{pmatrix}
\begin{Bmatrix}
1  & 1  & 1  \\
l' & s' & j' \\
l  & s  & j
\end{Bmatrix}
\langle n'l'|| r||n l \rangle
\langle s' || s ||s \rangle
,
\end{eqnarray}
where the reduced matrix elements of coordinate and spin operators are given by the equations
\begin{eqnarray}
    \langle n'l'|| r||n l \rangle = (-1)^{l'}\sqrt{(2l'+1)(2l+1)}
     \\\nonumber
 \times
    \begin{pmatrix}
        l' & 1 & l \\ 
        0  & 0 & 0
    \end{pmatrix}
    \int\limits_{0}^{\infty}dr r^3 R_{n'l'}(r)R_{nl}(r)
    ,
\end{eqnarray}
\begin{eqnarray}
    \langle s' || s ||s \rangle = \delta_{s's}\sqrt{s(s+1)(2s+1)}
 ,
\end{eqnarray}
respectively, and $R_{nl}(r)$ is a solution of radial part of Schr\"odinger equation for the hydrogen-like atom.

The matrix element of spin-orbit interaction can be evaluated in a similar manner as follows
\begin{eqnarray}
  \langle n'l's'j'm' | r^{-3}(\bm{l}\times \bm{s})_{1q} | nlsjm\rangle   = \qquad
\\\nonumber
 -
\mathrm{i}\sqrt{6} 
(-1)^{j'-m'}\sqrt{(2j'+1)(2j+1)} \times
\\\nonumber
\begin{pmatrix}
    j' & 1 & j \\
   -m' & q & m
\end{pmatrix}
\begin{Bmatrix}
1  & 1  & 1  \\
l' & s' & j' \\
l  & s  & j
\end{Bmatrix}
\langle n'l'|| r^{-3} ||n l \rangle
\langle s' || s ||s \rangle
,
\end{eqnarray}
where
\begin{eqnarray}
    \langle n'l'||r^{-3} ||n l \rangle =  \delta_{l'l}\sqrt{l(l+1)(2l+1)}
    \\\nonumber
    \times
    \int\limits_{0}^{\infty}dr r^2R_{n'l'}(r)\left(\frac{1}{r^3}\right)R_{nl}(r)
\end{eqnarray}
Likewise, we find
\begin{align}
    \label{LS}
&    
 \langle n'l's'j'm_{j'}|r^{-3}(\bm{l}\cdot\bm{s})| n l s j m_{j} \rangle = \delta_{l'l}\delta_{s's}\delta_{j'j}
  \\\nonumber
  \times
&  
 \delta_{m_{j'}m_{j}}(-1)^{j+l+s'}
 \begin{Bmatrix}
     l' & l & 1 \\
     s  & s' & j
 \end{Bmatrix}
  \sqrt{l(l+1)(2l+1)}
  \\\nonumber
  \times
&  
 \sqrt{s(s+1)(2s+1)}
% \\\nonumber
% \times
 \int\limits_{0}^{\infty}drr^2 R_{n'l'}(r)
\left(
\frac{1}{r^3}
\right)
R_{nl}(r)
.
\end{align}

The matrix element of components of position operator can be evaluated as follows
\begin{eqnarray}
    \langle n'l's'j'm'|r_{1q}|n lsjm \rangle =     \qquad
    \\\nonumber
    \delta_{s's}(-1)^{j+l'+s-1}
     (-1)^{j'-m'}
    \sqrt{(2j'+1)(2j+1)}    
        \\\nonumber
    \times      
    \begin{pmatrix}
        j' &1 & j\\
        -m' & q & m
    \end{pmatrix}
    \begin{Bmatrix}
        l & s & j \\
        j' &   1 & j
    \end{Bmatrix}  
    \langle n'l'||r ||n l \rangle
    .
\end{eqnarray}

% \begin{eqnarray}
%     \langle n'l's'j'm'|r^{-1}|n lsjm \rangle = \delta_{s's}(-1)^{j+l'+s-1}
%     \\\nonumber
%     \times
%     \sqrt{2j+1}
%     \times      
%     \begin{Bmatrix}
%         l & s & j \\
%         j' &   1 & j
%     \end{Bmatrix}  
%     \langle n'l'||r^{-1} ||n l \rangle
% \end{eqnarray}

% \begin{eqnarray}
%      \langle n'l'||r^{-1} ||n l \rangle = \sqrt{2l'+1}\int\limits_{0}^{\infty}drr^2 R_{n'l'}(r)
% \left(
% \frac{1}{r}
% \right)
% R_{nl}(r)
% .
% \end{eqnarray}

For the analytical calculation of the Lamb shift, we also need the commutation relation 
\begin{eqnarray}
    \label{PH}
    [p_{i},H] = - \mathrm{i}\nabla_{i} V
    ,
\end{eqnarray}
which is used in the proof of equality
\begin{align}
\label{off_xa9}
& 
\langle 
\phi_{n'} |
p_{i}
(H_{S}-E_{a}\pm k)
p_{i}
| \phi_{n} 
\rangle
\\\nonumber
& 
=
\frac{1}{2}
\langle 
\phi_{n'} |
\Delta V
| \phi_{n} 
\rangle 
+ \left(\frac{1}{2}(E_{n'} + E_{n})-E_{a} \pm k   \right)
\\\nonumber
&
\times
\langle \phi_{n'} | p^2 | \phi_{n} 
\rangle
.
\end{align}
For the case when $n'=n=a$ and $k=0$, Eq. (\ref{off_xa9}) turns to Eq. (\ref{xa9}) in the main text. 

In order to align the algebraic expressions of the divergent contributions in both the low-energy and high-energy regions, we also employ the following commutation relations:
\begin{eqnarray}
    \label{rp}
    [r_{i},p_{j}] = \mathrm{i}\delta_{ij}
    ,
\end{eqnarray}
\begin{eqnarray}
    \label{HHR}
    \langle n' | r_{i} | n \rangle (E_{n'}-E_{n})^2 = \langle n' |[H,[H,r_{i}]]| n\rangle
    .
\end{eqnarray}

% \begin{align*}
% [\alpha_{\mu}, \gamma_{\nu}] &= 
% \\\nonumber
% \alpha_{\mu}\gamma_{\nu} - \gamma_{\nu}\alpha_{\mu} \
% &= \frac{1}{2}(\gamma_{\mu}\gamma_{\nu} + \gamma_{\nu}\gamma_{\mu})\gamma_{\nu} - \frac{1}{2}(\gamma_{\nu}\gamma_{\mu} + \gamma_{\mu}\gamma_{\nu})\gamma_{\mu} \
% \\\nonumber
% &= \frac{1}{2}({\gamma_{\mu},\gamma_{\nu}}\gamma_{\nu} - {\gamma_{\nu},\gamma_{\mu}}\gamma_{\mu}) \
% &= \frac{1}{2}(2g_{\mu\nu}\mathbf{1} - 2g_{\nu\mu}\mathbf{1}) \
% &= i\sigma_{\mu\nu}
% \end{align*}

%merlin.mbs apsrev4-1.bst 2010-07-25 4.21a (PWD, AO, DPC) hacked
%Control: key (0)
%Control: author (8) initials jnrlst
%Control: editor formatted (1) identically to author
%Control: production of article title (-1) disabled
%Control: page (0) single
%Control: year (1) truncated
%Control: production of eprint (0) enabled
%

\end{document}